\documentclass[aps,superscriptaddress,amsmath,amssymb,floatfix,twocolumn,showpacs,amsfonts,longbibliography]{revtex4-2}
\usepackage{times}
\usepackage[varg]{txfonts}
\usepackage{textcomp}
\usepackage{graphicx}
\usepackage{subfigure}
\usepackage{color}
\usepackage[colorlinks=true,citecolor=blue,urlcolor=blue,linkcolor=blue,hyperindex]{hyperref}
\usepackage{braket}
\usepackage{overpic}
\usepackage{bm}

\def\non{\nonumber}

\begin{document}

\title{Topological magnons on the triangular kagome lattice}
\author{Meng-Han Zhang}
\affiliation{State Key Laboratory of Optoelectronic Materials and Technologies, Center for Neutron Science and Technology, Guangdong Provincial Key Laboratory of Magnetoelectric Physics and Devices, School of Physics, Sun Yat-Sen University, Guangzhou 510275, China}

\author{Dao-Xin Yao}
\email[Corresponding author:]{yaodaox@mail.sysu.edu.cn}
\affiliation{State Key Laboratory of Optoelectronic Materials and Technologies, Center for Neutron Science and Technology, Guangdong Provincial Key Laboratory of Magnetoelectric Physics and Devices, School of Physics, Sun Yat-Sen University, Guangzhou 510275, China}
\affiliation{International Quantum Academy, Shenzhen 518048, China}
\date{\today}

\begin{abstract}

We present the topology of magnons on the triangular kagome lattice (TKL) by calculating its Berry curvature, Chern number and edge states. In addition to the ferromagnetic state, the TKL hosts ferrimagnetic ground state as its two sublattices can couple with each other either ferromagnetically or antiferromagnetically. Using Holstein-Primakoff (HP) boson theory and Green's function approach, we find that the TKL has a rich topological band structure with added high Chern numbers compared with the kagome and honeycomb lattices. The magnon edge current allows a convenient calculation of thermal Hall coefficients and the orbital angular momentum gives correlation to the Einstein-de Haas effect. We apply the calculations to the TKL and derive the topological gyromagnetic ratio showing a nonzero Einstein-de Haas effect in the zero temperature limit. Our results render the TKL as a potential platform for quantum magnonics applications including high-precision mechanical sensors and information transmission.

\end{abstract}

\maketitle
%%%%%%%%%%%%%%%%%%%%%%%%%%%%%%%%%%%%%%%%%%%%%%%%%%%%%%%
\section{INTRODUCTION}\label{sec:intro}
%%%%%%%%%%%%%%%%%%%%%%%%%%%%%%%%%%%%%%%%%%%%%%%%%%%%%%%
The discovery of gyromagnetism~\cite{PhysRev.6.239}, the interconversion between spin and mechanical rotational motions, revealed that the origin of magnetism was the intrinsic angular momentum of electrons. By determining the gyromagnetic ratio~\cite{PhysRevB.79.104410}, the Einstein–de Haas (EdH) effect provides a more accurate measurement of the rotational motion rather than electron-spin resonance or ferromagnetic resonance~\cite{1915KNAB...18..696E}. Recent studies show that circularly polarized phonons can absorb the angular momentum of the spin system, which provide an atomistic picture of the EdH effect~\cite{PhysRevB.99.064428}. Indicating the transfer between a magnetic moment and a macroscopic mechanical rotation, the EdH technique attracts increasing attention and has important consequences in the fields of quantum thermal transport~\cite{Wang2008}, nano-magneto-mechanical systems~\cite{PhysRevLett.94.167201,PhysRevLett.104.027202,PhysRevB.95.134447,PhysRevB.75.014430}, spintronics~\cite{PhysRevLett.106.076601}, magnonics~\cite{PhysRevApplied.9.024029}, ultrafast magnetism~\cite{RevModPhys.90.015005,PhysRevLett.118.117203}.

As the bosonic analog of the electron system, the orbital motions of magnons are driven by the Berry curvature in momentum space from the topological band structure~\cite{PhysRevLett.117.227201,PhysRevLett.117.217203,PhysRevLett.117.217202,Owerre_2017}. These orbital motions cause the thermal Hall effect arising from the edge current of magnons. It has been observed experimentally in a number of three-dimensional ferromagnetic pyrochlores($Lu_2V_2O_7$, $Ho_2V_2O_7$, and $In_2Mn_2O_7$)~\cite{science.1188260,PhysRevB.85.134411}. According to the linear response theory, there is a reduced angular momentum generated by the orbital motion of the magnon~\cite{physrevb.84.184406,physrevlett.106.197202}. The reduced angular momentum per unit cell consists of two components, the edge current and the self-rotation, and is related to the EdH effect~\cite{physrevresearch.3.023248}. We apply the calculation of angular momentum on various lattices finding that the triangular kagome lattice (TKL) has a larger response than the kagome and honeycomb lattices both in the thermal Hall effect and the EdH effect. Our calculations are applicable to the magnon transport theory which makes a remarkable progress in coding and processing information~\cite{YUAN20221} due to the small dissipation significantly reducing the energy consumption~\cite{nphys3347,acs.nanolett.8b00492}.

Distinct from the ordinary bipartite lattices, the TKL with nine spins in the unit cell can produce magnetic long-range order in both ferro- and ferrimagnetic states~\cite{physrevb.78.224410,physrevb.78.024428,PhysRevB.78.024427,Stre_ka_2009,Owerre_2016}. It is worthwhile to study the topological properties of magnons and related effects for these ordered states on the TKL which has been realized experimentally in a two-dimensional metal organic framework halide series, $Cu_9X_2(cpa)_6$($X$=$F, Cl, Br$; cpa=anion of 2-carboxypentonic acid)~\cite{AIP.8.101404,AIP.10.025025}. It is best described as a spin frustrated TKL on a layered metal organic framework formed by inserting an extra set of triangles inside of the kagome triangles~\cite{physreve.98.012127,physrevb.77.134402}. With an odd number of spins in the unit cell, the TKL gives rise to three times the unit cell of the kagome lattice~\cite{PhysRevLett.115.147201}, and hence a new platform to explore topological magnon effects. The Dzyaloshinsky-Moriya(DM) interaction induces a fictitious magnetic flux and leads to the existence of nonzero Berry curvature. With different Heisenberg exchange couplings, the nonzero DM interaction on the TKL induces a rich phase diagram accompanied by the topologically protected gapless edge modes. As the inversion-symmetry-breaking can eliminate the degeneracy of energy bands, the TKL provides a promising avenue for realizing exotic quantum phenomena~\cite{PhysRevLett.108.246402,s11467-009-0075-x}, magnon thermal devices~\cite{physrevlett.104.066403,PhysRevLett.120.097702}, and magnon mechanical devices~\cite{physrevresearch.3.023248}.

In this work, we theoretically study the topological magnon excitations on the TKL proposing effective realizations for both the thermal Hall effect and the EdH effect. We track the corresponding DOS of edge states by using the real-space Green's function. The thermal Hall conductance behavior $\kappa^{xy}$ provides useful insights of the magnon transport as it can detect the charge-neutral quasiparticles that would not directly couple to electromagnetic probes. The EdH physics is properly captured by the thermal dependence of the gyromagnetic ratio as a function of the different material parameters. This behavior is inherited from the topology of the magnon bulk bands and further confirms the sign change behavior of thermal Hall response. Through estimating the gyromagnetism, we find that the TKL has a larger EdH response than the kagome and honeycomb lattices.

This paper is organized as follows. In Sec.~\ref{sec:model} we introduce the model (Sec.~\ref{subsec:TKSM}) and present the equations for spin-wave Hamiltonians using the spin-wave theory and HP boson theory. Chern number and thermal Hall conductance are defined in Sec.~\ref{subsec:CNTH}. Then, we present edge state geometry, the formalism of Green's function (Sec.~\ref{subsec:CEGF}) and angular momentum expressions (Sec.~\ref{subsec:AMGR}). In Sec.~\ref{sec:results} we present our results on topological energy bands (Sec.~\ref{subsec:teb}), density of states (Sec.~\ref{subsec:dos}), thermal Hall effect (Sec.~\ref{subsec:mhe}) and finally we discuss the Einstein-de Haas effect of our results (Sec.~\ref{subsec:EdH}). In Sec.~\ref{sec:conclu} we discuss and conclude our findings.
%%%%%%%%%%%%%%%%%%%%%%%%%%%%%%%%%%%%%%%%%%%%%%%%%%%%%%%%%%%%%%%%
\section{MODEL AND METHODS}\label{sec:model}
%%%%%%%%%%%%%%%%%%%%%%%%%%%%%%%%%%%%%%%%%%%%%%%%%%%%%%%%%%%%%%%%
%%%%%%%%%%%%%%%%%%%%%%%%%%%%%%%%%%%%%%%%%%%%%%%%%%%%%%%%%%%%%%%%
\subsection{Triangular Kagome Spin Model}\label{subsec:TKSM}
%%%%%%%%%%%%%%%%%%%%%%%%%%%%%%%%%%%%%%%%%%%%%%%%%%%%%%%%%%%%%%%%
\begin{figure}[t]
\centering
{
\subfigure[]{
\includegraphics[width=1.5in]{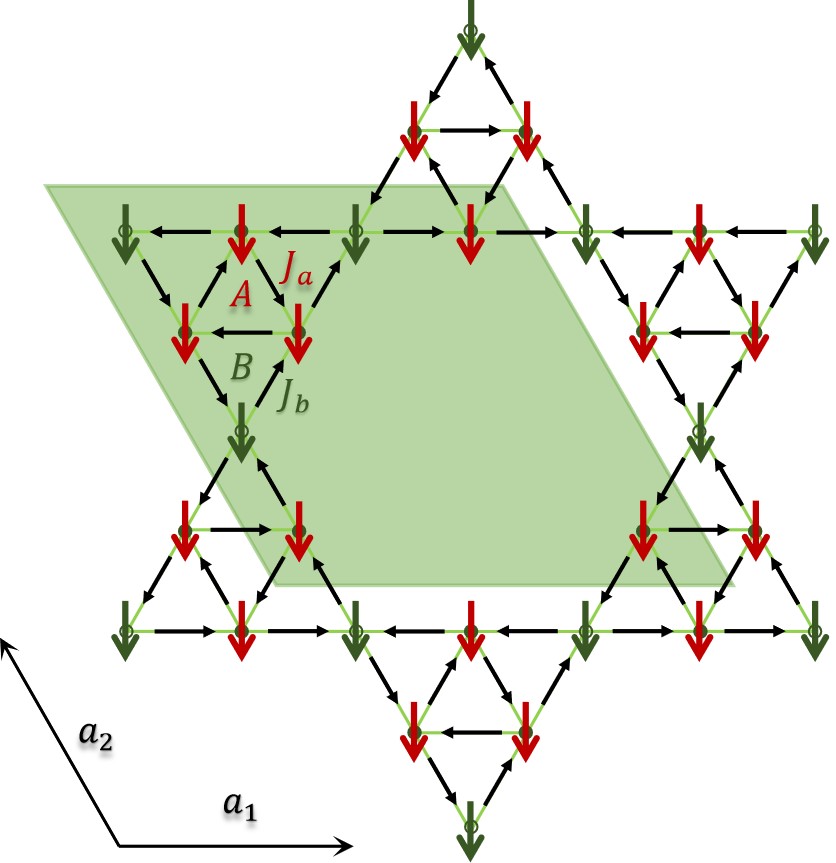}\label{fig1a}}
}
{
\subfigure[]{
\includegraphics[width=1.5in]{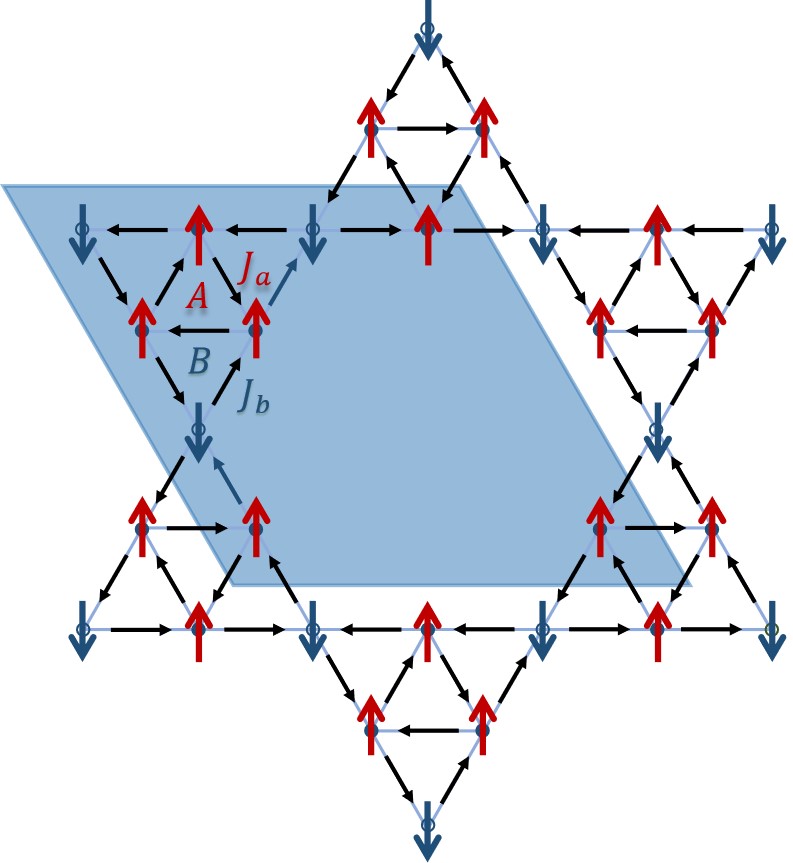}\label{fig1b}}
}
\caption{Schematics of the TKL with shaded regions that represent the unit cell. $a_{1}$, $a_{2}$ are basis vectors of the primitive unit cell and the arrows within the single triangular blocks indicate the configurations of the DM-induced flux, highlighted by black solid arrows. The red arrows are A sites and others are B sites. (a) Ferromagnetic ground state with $J_{a}$$>$0 and $J_{b}$$>$0. (b) Ferrimagnetic ground state with $J_{a}$$>$0 and $J_{b}$$<$0.}
\label{Fig1}
\end{figure}
To present the method of approach in a concrete background we consider a Heisenberg model on the TKL with nine spins in the unit cell, where the total Hamiltonian is given by
\begin{align}
\mathcal{H}=\mathcal{H}_0 +\mathcal{H}_{\mathrm{DM}} +\mathcal{H}_{K} +\mathcal{H}_B,
\end{align}
Our model Hamiltonian contains the nearest-neighbor Heisenberg exchange interactions, where the $\mathcal{H}_0$ is
\begin{align}
\mathcal{H}_0=-J_a\sum_{\langle mn\rangle} \boldsymbol{S}_m \cdot \boldsymbol{S}_n -J_b\sum_{\langle mn\rangle} \boldsymbol{S}_m \cdot \boldsymbol{S}_n,
\end{align}
and $J_a$, $J_b$ are two types of the nearest-neighbor exchange couplings within the sublattice $\Delta$ ($A$-trimers indicated with red sites) and $\nabla$ ($B$-trimers indicated with green sites) as shown in Fig.~\ref{Fig1}. The $\mathcal{H}_{\mathrm{DM}}$ term represents the nearest-neighbor DM interaction which is usually dominant perturbative anisotropy to the Heisenberg exchange interactions. Therefore, it could be considered as
\begin{align}
\mathcal{H}_\mathrm{DM}^{} = \sum_{\langle mn\rangle} \boldsymbol{D}_{mn}^{} \cdot (\boldsymbol{S}_m^{} \times \boldsymbol{S}_n^{}).
\end{align}
Here we introduce the anisotropy term and the Zeeman term to have the magnetic order even at finite temperature based on the Mermin-Wagner theorem~\cite{Torres_2014,PhysRevLett.17.1133}. The anisotropy term is given by
\begin{align}
\mathcal{H}_{K}=-K\sum_{\langle m\rangle} (S_m^z)^2,
\end{align}
where $K$ is the easy-axis anisotropy along the $z$-axis. And the external Zeeman magnetic field term is given by
\begin{align}
\mathcal{H}_B=-h\sum_{\langle m\rangle} S_m^z,
\end{align}
where $h =g \mu_B B$, $B$ is the external magnetic field.

The TKL has a ferromagnetic ground state for $J_{a}$$>$0 and $J_{b}$$>$0 in Fig.~\ref{fig1a}, while a ferrimagnetic ground state for $J_{a}$$>$0 and $J_{b}$$<$0 is shown in Fig.~\ref{fig1b}. Here we use the Holstein-Primakoff (HP) representation to study the magnetic excitations for the ordered states. The original spin Hamiltonian can be mapped to a bosonic tight binding model following the HP transformation:
\begin{align}
S^+_m&=S^x_m+iS^y_m=\sqrt{2S-\alpha^{\dag}_m\alpha_m}\alpha^{}_m,\non\\
S^-_m&=S^x_m-iS^y_m=\alpha^{\dag}_m\sqrt{2S-\alpha^{\dag}_m\alpha_m},\non\\
S^z_m&=S-\alpha^{\dag}_m\alpha^{}_m,
\end{align}
where $\alpha^{\dag}_m(\alpha_m^{})$ is the bosonic magnon creation (annihilation) operator at site $m$. Within the approximation of $\sqrt{2S-\alpha^{\dag}_m\alpha_m}$ $\rightarrow$$\sqrt{2S}$, the Hamiltonian has the form
\begin{align}
\mathcal{H}&=-\Big[\sum_{\langle mn\rangle_a}(J_a+i\nu_{mn}D)S\alpha^{\dag}_m\alpha^{}_n+\sum_{\langle mn\rangle_b}(J_b+i\nu_{mn}D)S\alpha^{\dag}_m\alpha^{}_n\non\\
&+H.c.\Big] +(2K+h)\sum_{\langle m\rangle}\alpha^{\dag}_m\alpha^{}_m+E^{}_0,
\end{align}
where $D$ is the $z$-component of the nearest-neighbor DM interaction, $E_0$ is ground state energy and $\nu_{mn}=\pm1$ corresponding to the direction of DM interaction. Subsequently, we perform the Fourier transformation using the definition
\begin{align}
\alpha^{\dag}_{\boldsymbol{k}}=\frac{1}{\sqrt{N}}\sum_me^{i\boldsymbol{k}\cdot\boldsymbol{R}_m}\alpha^{\dag}_m.
\end{align}
Thus, in the reciprocal space the Hamiltonian is given by
\begin{align}
\mathcal{H}=\sum_{\boldsymbol{k}}\psi^{\dag}_{\boldsymbol{k}}H(\boldsymbol{k})\psi^{}_{\boldsymbol{k}},
\end{align}
where $\psi^{\dag}_{\boldsymbol{k}}=(\alpha^{\dag}_{1,\boldsymbol{k}},\alpha^{\dag}_{2,\boldsymbol{k}},\alpha^{\dag}_{3,\boldsymbol{k}},\alpha^{\dag}_{4,\boldsymbol{k}},\alpha^{\dag}_{5,\boldsymbol{k}},\alpha^{\dag}_{6,\boldsymbol{k}},\alpha^{\dag}_{7,\boldsymbol{k}},\alpha^{\dag}_{8,\boldsymbol{k}},\alpha^{\dag}_{9,\boldsymbol{k}})$. The spin wave Hamiltonian matrix is
\begin{align}
S\left[
\begin{array}{ccc}
E_1 I_{3\times3} & A_{\boldsymbol{k}} & B_{\boldsymbol{k}} \\
A_{\boldsymbol{k}}^{\dag} & C_{\boldsymbol{k}} & 0_{3\times3} \\
B_{\boldsymbol{k}}^{\dag} & 0_{3\times3} & D_{\boldsymbol{k}} \\
\end{array}
\right],
\end{align}
with matrix $A_{\boldsymbol{k}}$ is
\begin{align}
\left[
\begin{array}{ccc}
-\gamma_1e^{-i\boldsymbol{k\cdot a_1}} & -\gamma_1e^{-i\boldsymbol{k\cdot a_2}} & 0 \\
-\gamma_1e^{i\boldsymbol{k\cdot a_1}} & 0 & -\gamma_1e^{i\boldsymbol{k\cdot(a_1-a_2)}} \\
0 & -\gamma_1e^{i\boldsymbol{k\cdot a_2}} & -\gamma_1e^{i\boldsymbol{k\cdot(-a_1+a_2)}} \\
\end{array}
\right],
\end{align}
matrix $B_{\boldsymbol{k}}$ is
\begin{align}
\left[
\begin{array}{ccc}
-\gamma_1e^{i\boldsymbol{k\cdot a_1}} & -\gamma_1e^{i\boldsymbol{k\cdot a_2}} & 0 \\
-\gamma_1e^{-i\boldsymbol{k\cdot a_1}} & 0 & -\gamma_1e^{i\boldsymbol{k\cdot(-a_1+a_2)}} \\
0 & -\gamma_1e^{-i\boldsymbol{k\cdot a_2}} & -\gamma_1e^{i\boldsymbol{k\cdot(a_1-a_2)}} \\
\end{array}
\right],
\end{align}
matrix $C_{\boldsymbol{k}}$ is
\begin{align}
\left[
\begin{array}{ccc}
E_2 & -\gamma_2e^{i\boldsymbol{k\cdot(a_1-a_2)}} & -\gamma_2e^{-i\boldsymbol{k\cdot a_2}} \\
-\gamma_2e^{i\boldsymbol{k\cdot(-a_1+a_2)}} & E_2 & -\gamma_2e^{-i\boldsymbol{k\cdot a_1}} \\
-\gamma_2e^{i\boldsymbol{k\cdot a_2}} & -\gamma_2e^{i\boldsymbol{k\cdot a_1}} & E_2 \\
\end{array}
\right],
\end{align}
and matrix $D_{\boldsymbol{k}}$ is
\begin{align}
\left[
\begin{array}{ccc}
E_2 & -\gamma_2e^{i\boldsymbol{k\cdot(-a_1+a_2)}} & -\gamma_2e^{i\boldsymbol{k\cdot a_2}} \\
-\gamma_2e^{i\boldsymbol{k\cdot(a_1-a_2)}} & E_2 & -\gamma_2e^{i\boldsymbol{k\cdot a_1}} \\
-\gamma_2e^{-i\boldsymbol{k\cdot a_2}} & -\gamma_2e^{-i\boldsymbol{k\cdot a_1}} & E_2 \\
\end{array}
\right],
\end{align}
where $E_1=4J_b+2K+h$, $E_2=2J_a+2J_b+2K+h$, $\gamma_1=J_b+i\nu_{mn}D$ and $\gamma_2=J_a+i\nu_{mn}D$. The lattice vectors are given by $\boldsymbol{a_1}=\frac{1}{4}(1, 0)a$ and $\boldsymbol{a_2}=\frac{1}{8}(-1, \sqrt{3})a$ with the lattice constant chosen as $a$=0.1nm. The energy bands obtained via diagonalizing the bilinear spin wave Hamiltonian are shown in Fig.~\ref{Fig2}.
\begin{figure}[t]
\centering
{
\includegraphics[width=3.3in]{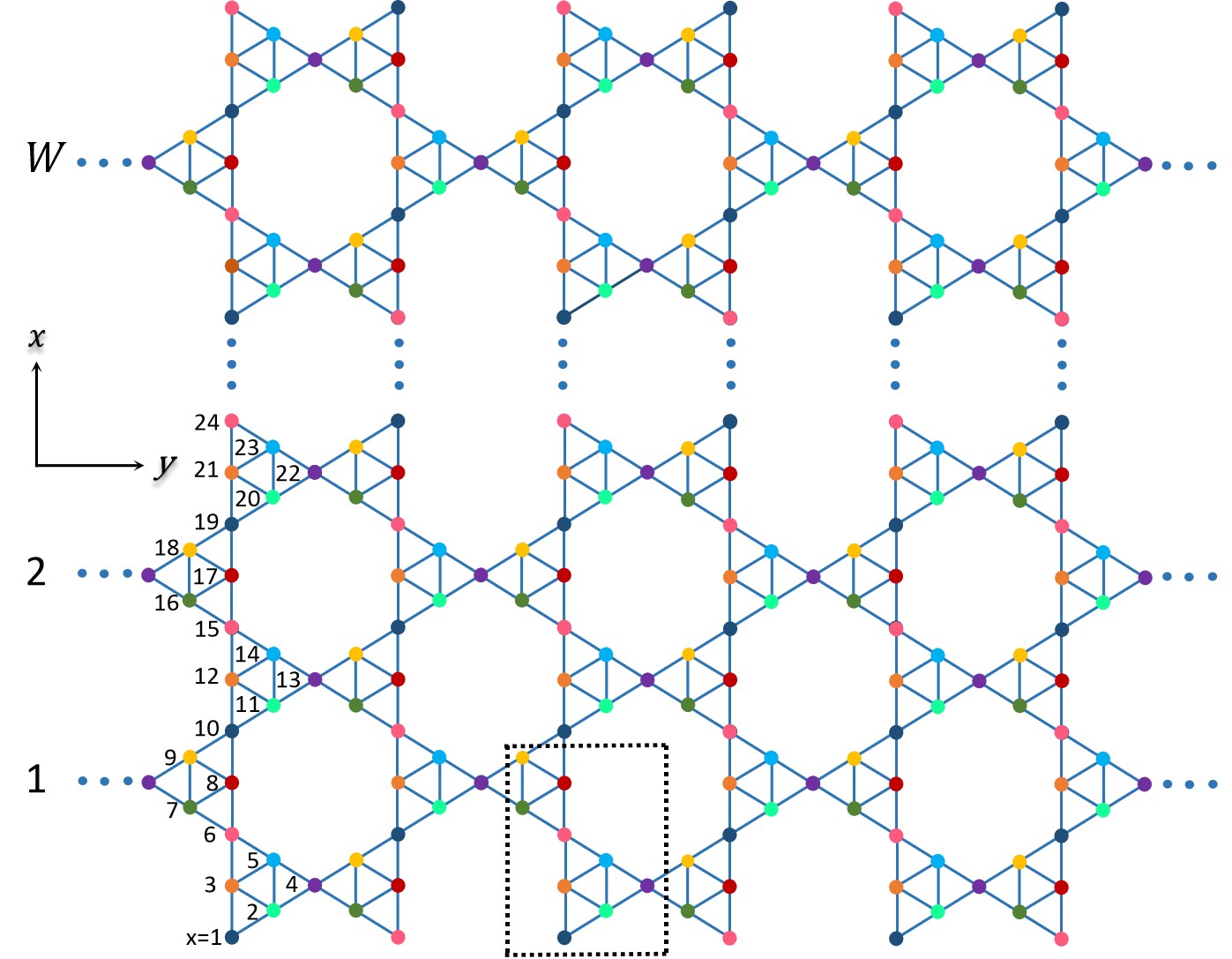}\label{fig13}
}
\caption{The TKL ribbon with periodic boundary condition along $y$-axis and open boundary condition along $x$-axis. The ribbon has W periodic one-dimensional chains, the numbers nearing sites are $x$ indices.}
\label{Fig13}
\end{figure}
%%%%%%%%%%%%%%%%%%%%%%%%%%%%%%%%%%%%%%%%%%%%%%%%%%%%%%%%%%%%%%%%
\subsection{Green's Functions in a Ribbon Sample}\label{subsec:CEGF}
%%%%%%%%%%%%%%%%%%%%%%%%%%%%%%%%%%%%%%%%%%%%%%%%%%%%%%%%%%%%%%%%
\begin{figure*}[t]
\centering
{
\subfigure[]{
\includegraphics[width=1.2in]{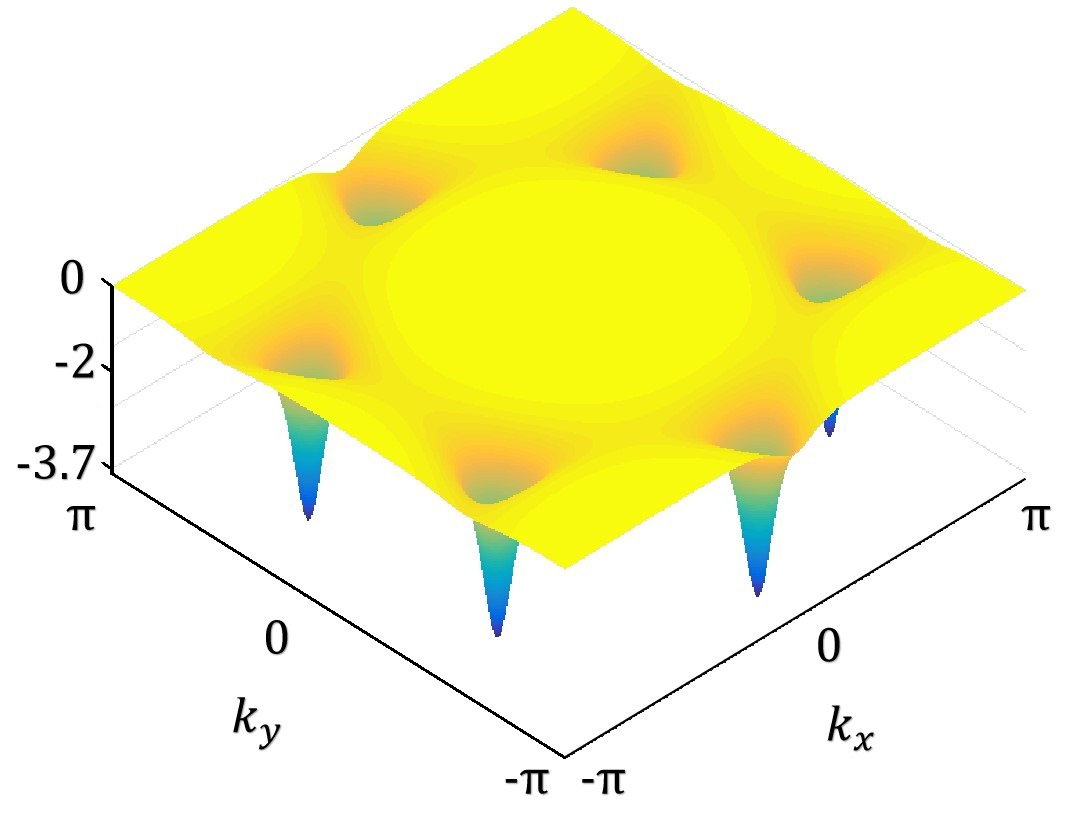}\label{fig22a}}
}
{
\subfigure[]{
\includegraphics[width=1.2in]{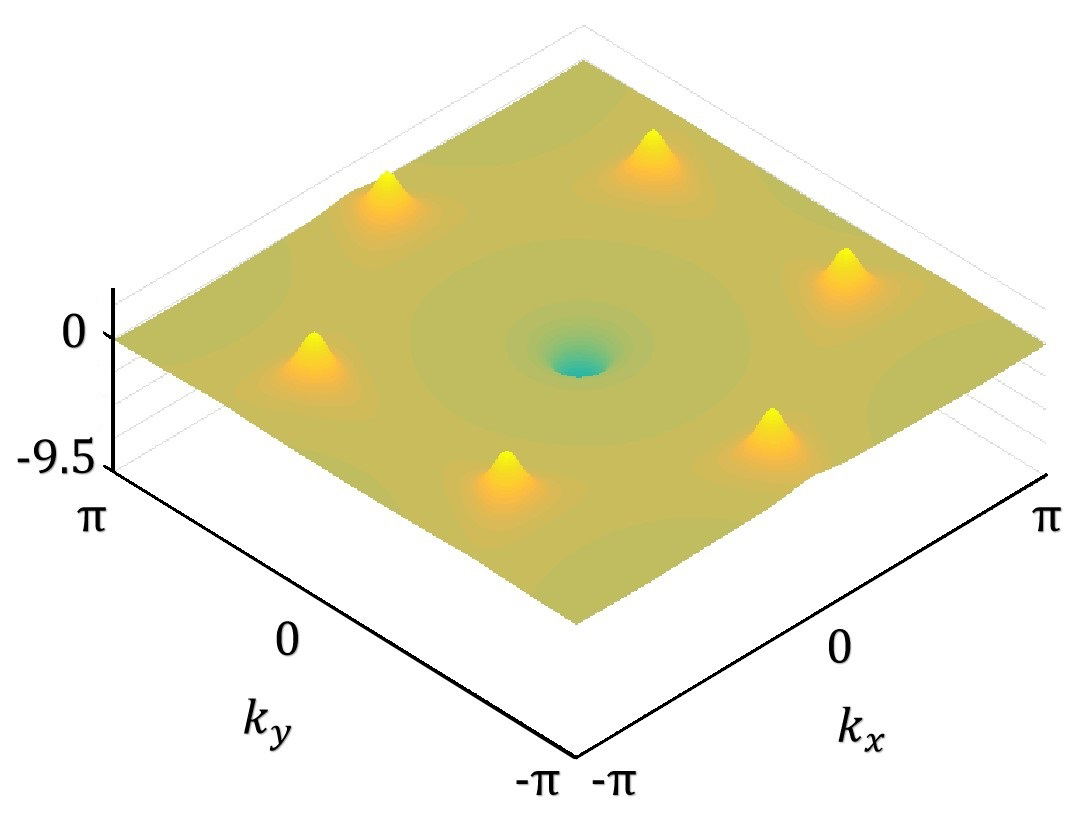}\label{fig22b}}
}
{
\subfigure[]{
\includegraphics[width=1.2in]{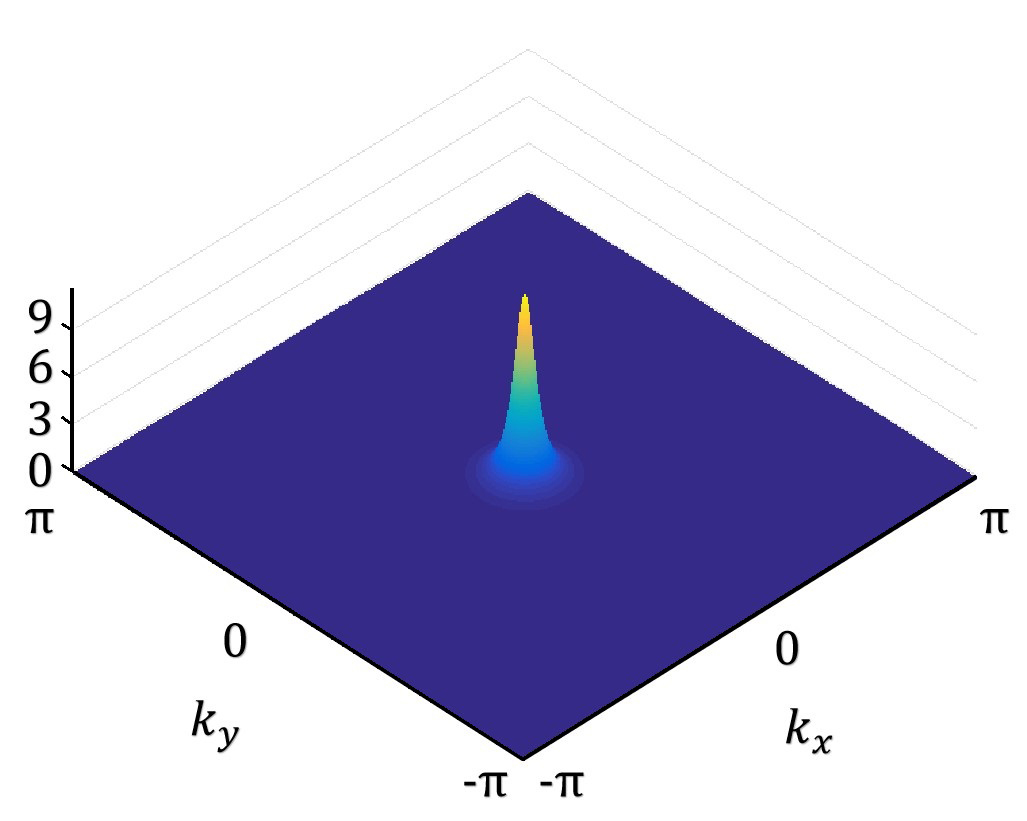}\label{fig22c}}
}
{
\subfigure[]{
\includegraphics[width=1.2in]{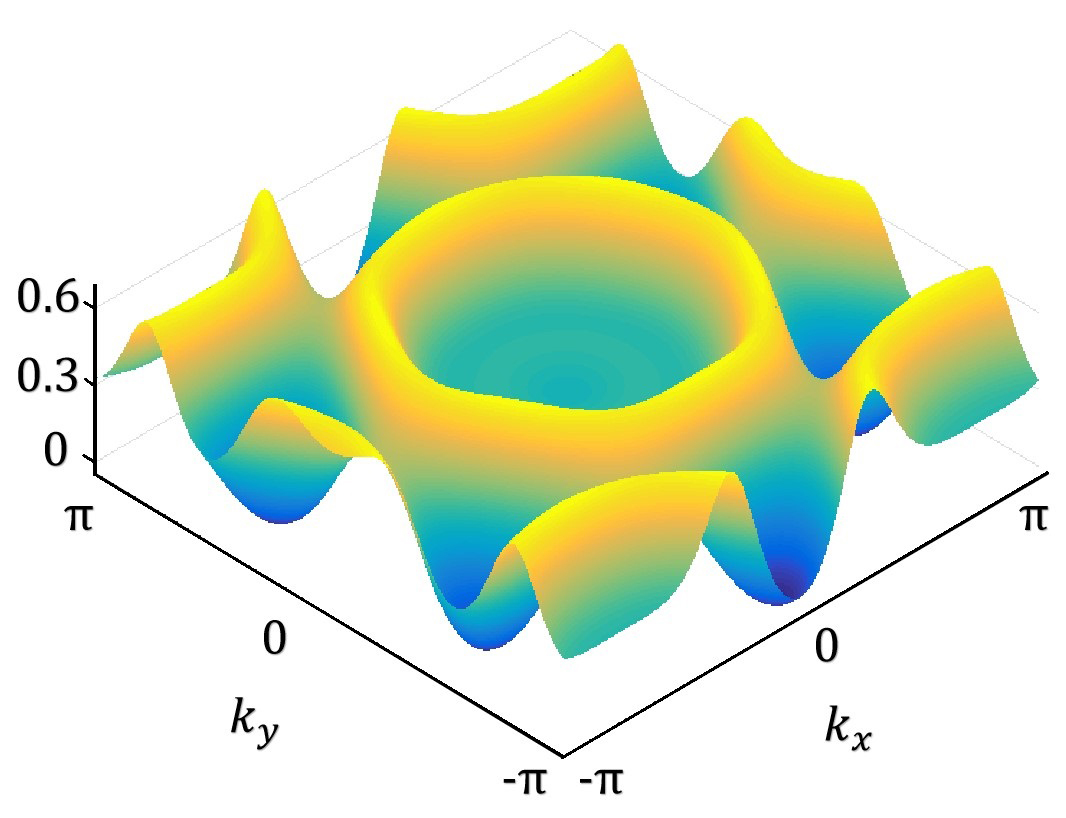}\label{fig22d}}
}
{
\subfigure[]{
\includegraphics[width=1.2in]{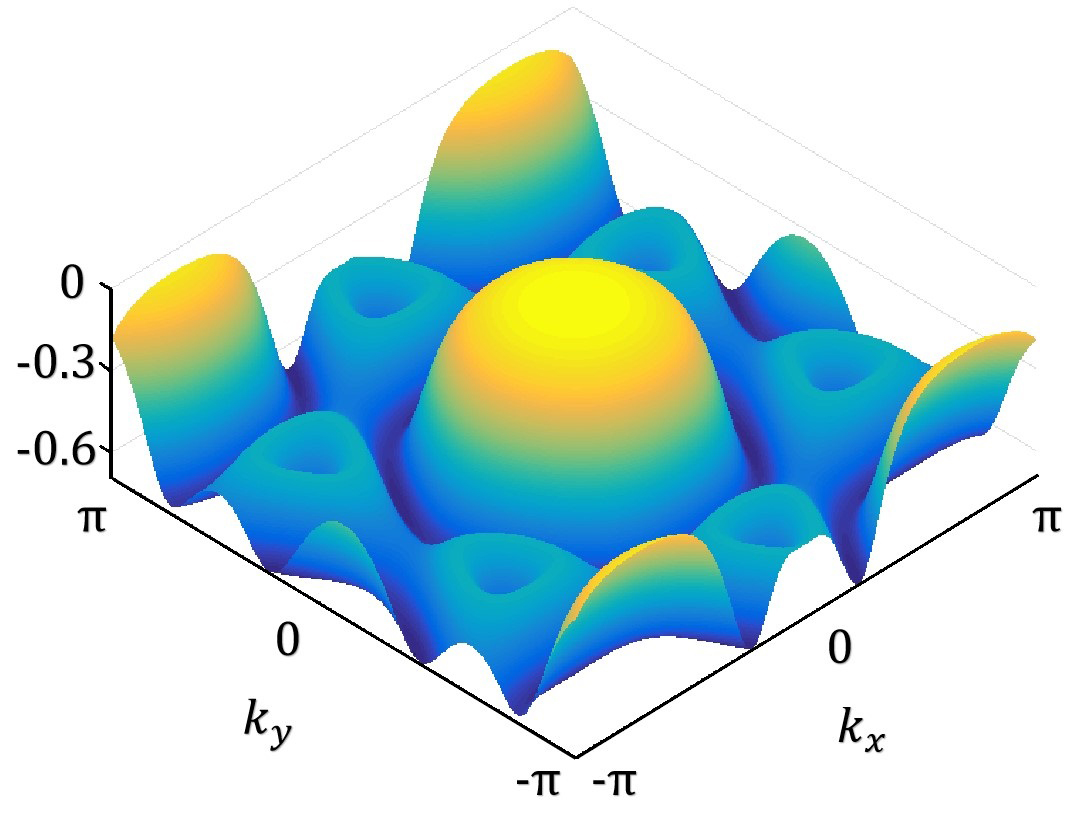}\label{fig22e}}
}
{
\subfigure[]{
\includegraphics[width=1.5in]{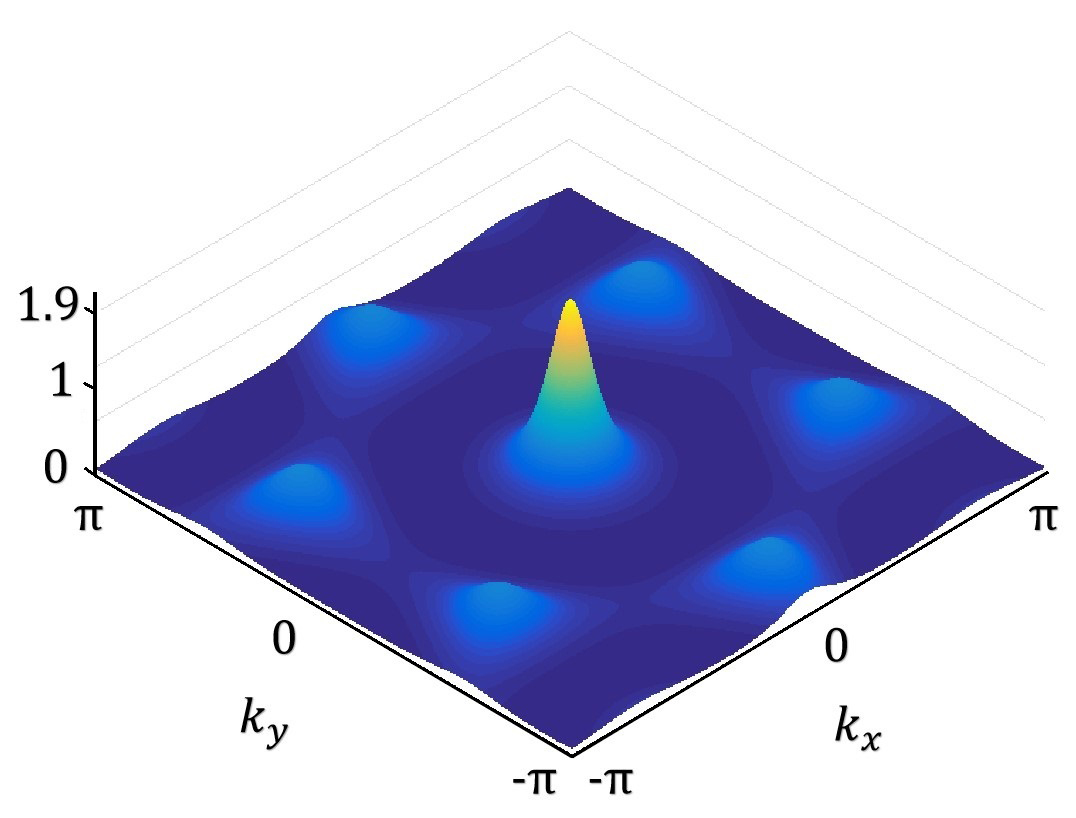}\label{fig22f}}
}
{
\subfigure[]{
\includegraphics[width=1.5in]{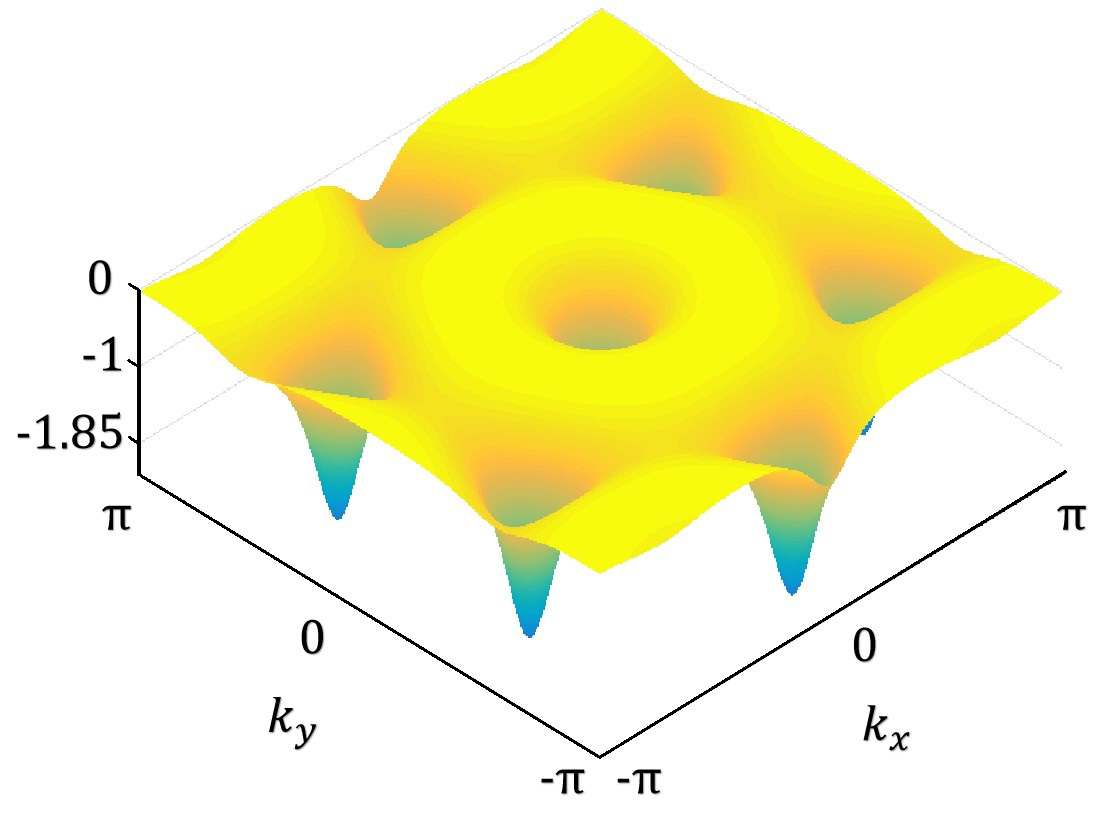}\label{fig22g}}
}
{
\subfigure[]{
\includegraphics[width=1.5in]{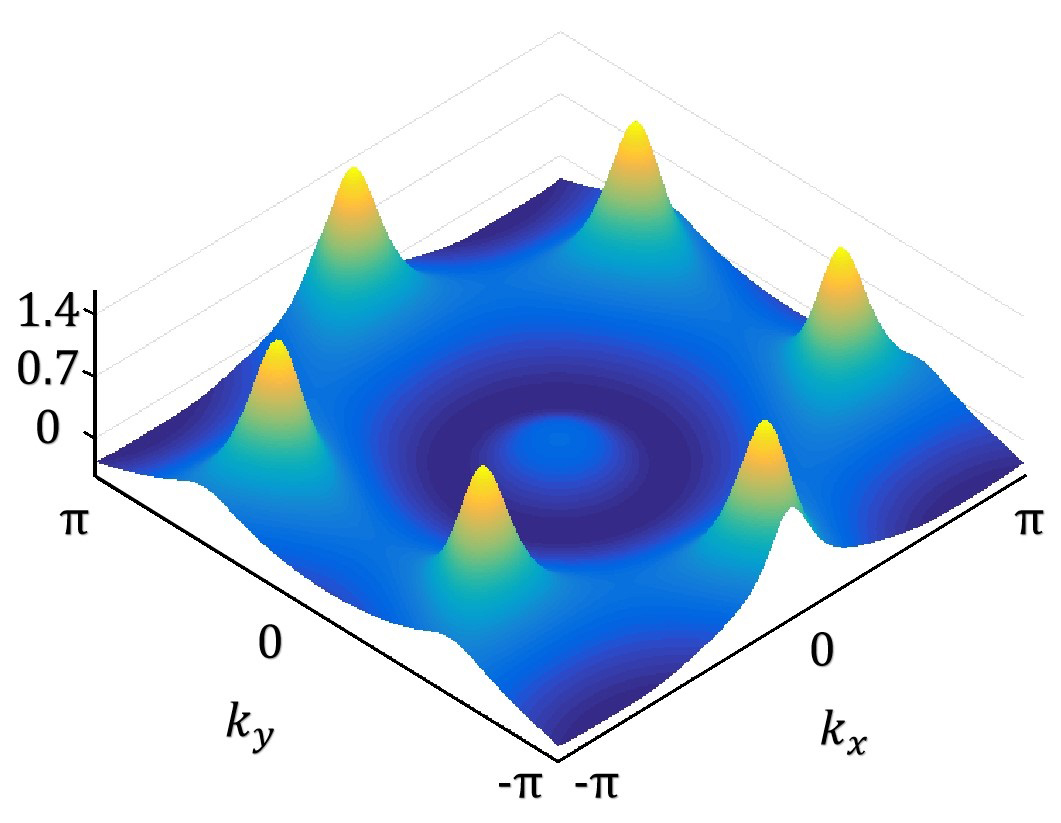}\label{fig22h}}
}
{
\subfigure[]{
\includegraphics[width=1.5in]{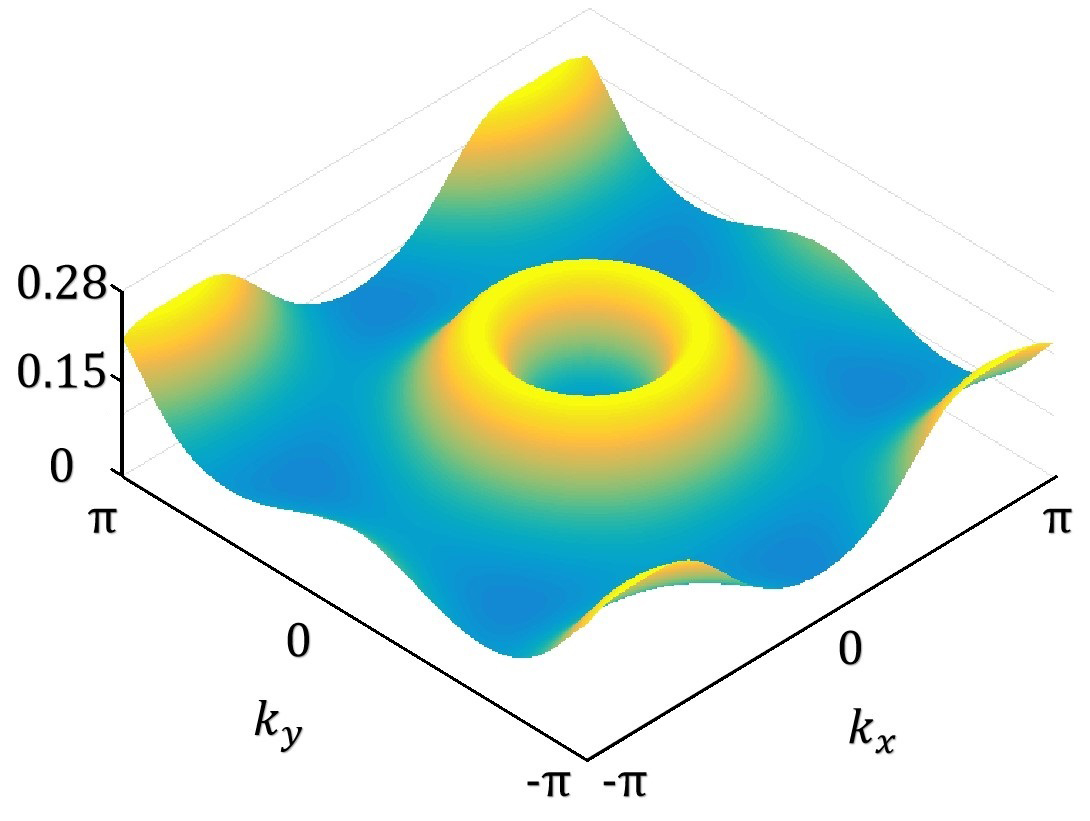}\label{fig22h}}
}
\caption{Berry curvature of magnon bands with $J_{a}$=0.5, $J_{b}$=1 and $D$=0.3. The (a), (b), (c), (d), (e), (f), (g), (h), (i) figures correspond to the first, second, third, fourth, fifth, sixth, seventh, eighth, ninth band (from lower to higher), respectively. The Chern numbers are given by $\{1, 0, 0, 2, -2, -1, 0, 1, -1\}$.}
\label{Fig22}
\end{figure*}
For a non-trivial topology of the bulk band structure, the edge states of the TKL appear in the DM-induced gaps for this ribbon sample. Due to the bulk-edge correspondence, the topological chiral gapless edge modes are related to the nonzero Chern numbers. We rewrite the Hamiltonian in the $(x, k_y)$ space as our ribbon sample is expanded to an open boundary condition along the $x$ direction and a periodic boundary condition along the $y$ direction.
\begin{align}
\alpha^{\dag}_{kx}=\frac{1}{\sqrt{N_{y}}}\sum_{m}e^{ik\boldsymbol{R}_{m}\cdot\boldsymbol{e}_{y}}\alpha^{\dag}_{mx},
\end{align}
where $x$ can run from $i_1$ to 9($W$-1)+$i_1$ ($i_1$=$\{$1, 2, 3, 4, 5, 6, 7, 8, 9$\}$) and $W$ denotes the number of periodic 1D chains along the $x$ direction. We replace $k_y$ by $k$. The formalism for calculating the band structure of the ribbon geometry is a $9W\times9W$ matrix-form Hamiltonian which is given by
\begin{align}
\mathcal{H}=\sum_{k}\varphi^{\dag}_{k}H(k)\varphi^{}_{k},
\end{align}
where $\varphi^{\dag}_{k}=(\alpha^{\dag}_{i_{1},k}, \alpha^{\dag}_{i_{1}+1,k},..., \alpha^{\dag}_{9(W-1)+i_{1},k})$ in the open boundary condition $\alpha^{\dag}_{0,k}|$0$\rangle$=$\alpha^{\dag}_{9W+1,k}|$0$\rangle$=0. The Hamiltonian matrix can be written as
\begin{align}
H(k)=
\left[
\begin{array}{ccccc}
G(k) & F(k)^{\dag} & 0 & \cdots & 0\\
F(k) & G(k) & F(k)^{\dag} & \ddots & \vdots \\
0 & F(k) & \ddots & \ddots & 0 \\
\vdots & \ddots & \ddots & \ddots & F(k)^{\dag} \\
0 & \cdots & 0 & F(k) & G(k) \\
\end{array}
\right],
\end{align}
where $G(k)$ and $F(k)$ are 9$\times$9 matrices with $G(k)_{ii}$=$E_0$$  (i$=$\{$1, 2, 3$\})$, $G(k)_{ii}$=$E_1$$ (i$=$\{$4, 5, 6, 7, 8, 9$\}$), $G(k)_{ij}$=$G(k)^{\dag}_{ji}$, $G(k)_{14}$ = $G(k)_{27}$ = $-\gamma_1e^{-ika_3}$, $G(k)_{15}$ = $G(k)_{29}$ = $G(k)_{38}$ = $F(k)_{36}$ = $-\gamma_1e^{-\frac{1}{2}ika_3}$, $G(k)_{17}$ = $G(k)_{24}$ = $-\gamma_1e^{ika_3}$, $G(k)_{18}$ = $G(k)_{26}$=$G(k)_{39}$ = $F(k)_{35}$ = $-\gamma_1e^{\frac{1}{2}ika_3}$, $G(k)_{45}$ = $G(k)_{79}$ = $-\gamma_2e^{\frac{1}{2}ika_3}$, $G(k)_{46}$ = $G(k)_{78}$ = $-\gamma_2e^{-\frac{1}{2}ika_3}$, $G(k)_{56}$ = $-\gamma_2e^{-ika_3}$, $G(k)_{89}$ = $-\gamma_2e^{ika_3}$, $G(k)_{ij}$ = 0 (otherwise), $F(k)_{ij}$ = 0 (otherwise), $a_3$ = 0.25$a$. We choose $W$=$20$ to ensure that the results are convergent with $W$. There are mainly two types of edges for the TKL: the zigzag edge and the armchair edge. In our case, we choose the armchair edge because the high symmetry points K and K' in the Brillouin zone overlap with each other along the $k_y$ direction~\cite{PhysRevB.48.11851}. Thus, the top and bottom edges are perpendicular to the $x$ direction shown in Fig.~\ref{Fig13}.

For the purpose of calculating transport properties of magnons, we introduce the retarded and advanced Green's functions.
\begin{align}
G^R(r, r')=\sum_{k,n} \frac{\alpha^{\dag}_{k,n}(r')\alpha^{}_{k,n}(r)}{\varepsilon+i\eta-H},G^A(r, r')=[G^R(r, r')]^{\dag},
\end{align}
where $\eta$ is a positive infinitesimal, $\varepsilon$ is the excitation energy, $r$ and $r'$ represent excitation and response respectively.
The spectral representation of the Green's function can be written as~\cite{pnas.1810003115}
\begin{align}
A=\sum_{k,n} \alpha^{}_{k,n}(r)\alpha^{\dag}_{k,n}(r')\frac{2\eta}{(\varepsilon-H)^2+\eta^2}.
\end{align}
And the DOS can also be defined as
\begin{align}
\mathbf{\rho}(\varepsilon) = \sum_{k,n} \alpha^{}_{k,n} \alpha^{\dag}_{k,n} \delta(\varepsilon-H) =\frac{\hbar \textmd{Tr}(A)}{2 \pi}.
\end{align}
With the above Green's functions, we can calculate the spectral function and the DOS of this ribbon sample. Both of them reflect the magnetic and topological properties of the TKL, which can solidify our proposal for the thermal Hall effect and the EdH effect.
%%%%%%%%%%%%%%%%%%%%%%%%%%%%%%%%%%%%%%%%%%%%%%%%%%%%%%%%%%%%%%%%
\subsection{Berry Curvature and Thermal Hall Conductance}\label{subsec:CNTH}
%%%%%%%%%%%%%%%%%%%%%%%%%%%%%%%%%%%%%%%%%%%%%%%%%%%%%%%%%%%%%%%%
In our model, nontrivial band topology can be characterized by a nonzero Berry curvature defined via the eigenstates of the system~\cite{RevModPhys.82.3045}. And a nontrivial band topology arises only when the system exhibits the nontrivial gap and edge state modes in the spin wave excitation spectra. In the case of two dimensional non-interacting magnons, generally topological invariant like Chern number denotes the topological nature of reciprocal space. We calculate the Berry connection in the reciprocal space of the TKL as
\begin{align}
A_{n}^{\lambda}=i\langle\psi_{\lambda}|\nabla_{\boldsymbol{k}_{n}}|\psi_{\lambda}\rangle.
\end{align}
with $|\psi_{\lambda}\rangle$ being a normalized wave function of the $\lambda$th Bloch band such that $H(\boldsymbol{k})|\psi_{\lambda}\rangle=E_{\lambda}(\boldsymbol{k})|\psi_{\lambda}\rangle$. The Berry connection is not a gauge invariant quantity but the Berry curvature is gauge invariant. The form of Berry curvature is given by
\begin{align}
\Omega_{\lambda \boldsymbol{k}}=i \sum_{n \neq \lambda}\frac{[{\langle\ \lambda\left| {\nabla_{\boldsymbol{k}} H(\boldsymbol{k})}\right| n}\rangle \times {\langle\ n \left| {\nabla_{\boldsymbol{k}} H(\boldsymbol{k})}\right| \lambda}\rangle]_z }{(E_\lambda-E_n)^2},
\end{align}

The associated Chern number assigned to the $n$th band is defined by
\begin{align}
C_{n} = \frac{1}{2\pi} \int_{BZ} d^2 k \Omega_{n \boldsymbol{k}}.
\end{align}

The Chern number is always a quantized integer in the Brillouin zone. When the gap between two bands is finite but very small, in general the Berry curvature is mostly concentrated around the point of direct gap between the bands. We show the Berry curvature of magnon bands in Fig.~\ref{Fig22} with $D$=0.3, $J_{a}$=0.5, and $J_{b}$=1.

Being charge neutral particles, magnons are not affected by external electric field and conventional electric field driven Hall effect can not be observed directly. Based on the semiclassical theory, the thermal gradient along topological magnon system would drive a transverse magnon current known as the thermal Hall effect. In our TKL system, the transverse current is understood as a consequence of the presence of chiral edge states induced by the DM interaction. We calculate the thermal Hall conductivity $\kappa_{xy}$ by the Kubo formula. It can be expressed as a weighted summation of the Berry curvature~\cite{physrevb.84.184406,PhysRevLett.49.405}
\begin{align}
\kappa_{xy}=-\frac{k_B^2 T}{4\pi^2 \hbar a} \sum_{n,\boldsymbol{k}} c_2[\rho(\varepsilon_{n \boldsymbol{k}})] \Omega_{n \boldsymbol{k}},
\end{align}
where $k_B$ is the Boltzmann constant, $T$ is the temperature and $\rho(\varepsilon_{n \boldsymbol{k}})$ $=[e^{\varepsilon_{n \boldsymbol{k}}/k_{B}T}-1]^{-1}$ is the Bose function. We choose the lattice constant $a$=0.1nm as the typical layer spacing for practical calculation. The $c_{2}(x)$ is defined as
\begin{align}
c_2=(1+x)(\ln\frac{1+x}{x})^2-(\ln x)^2-2Li_2(-x),
\end{align}
where $Li_2(x)$ is the polylogarithmic function. Considering the thermal fluctuation, we calculate the deviation of sublattice magnetization from the saturation value.
\begin{align}
\Delta m=S-\langle S^z_m\rangle = \langle\alpha^{\dag}_m\alpha_m\rangle=\sum_{n,\boldsymbol{k}} \rho(\varepsilon_{n \boldsymbol{k}}),
\end{align}
where the Curie temperature $T_c$ is determined by $\Delta m$ ($T_c$)=$S$.

\begin{figure*}[t]
\centering
{
\subfigure[]{
\includegraphics[width=1in]{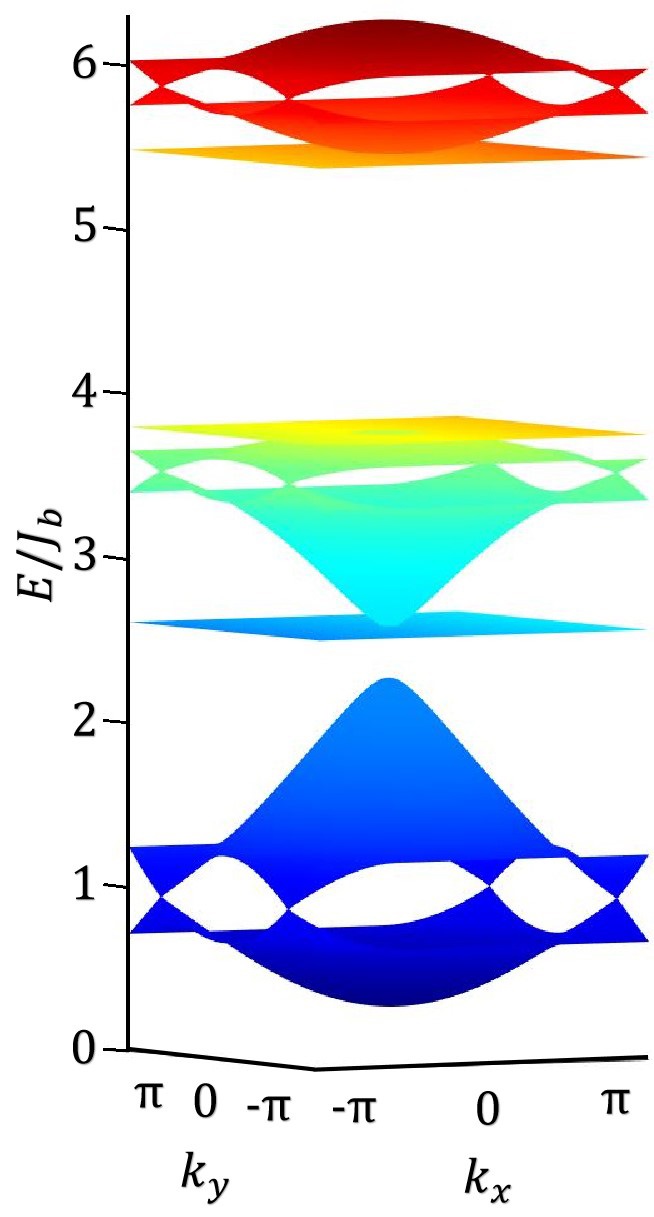}\label{fig2a}}
}
{
\subfigure[]{
\includegraphics[width=1.0in]{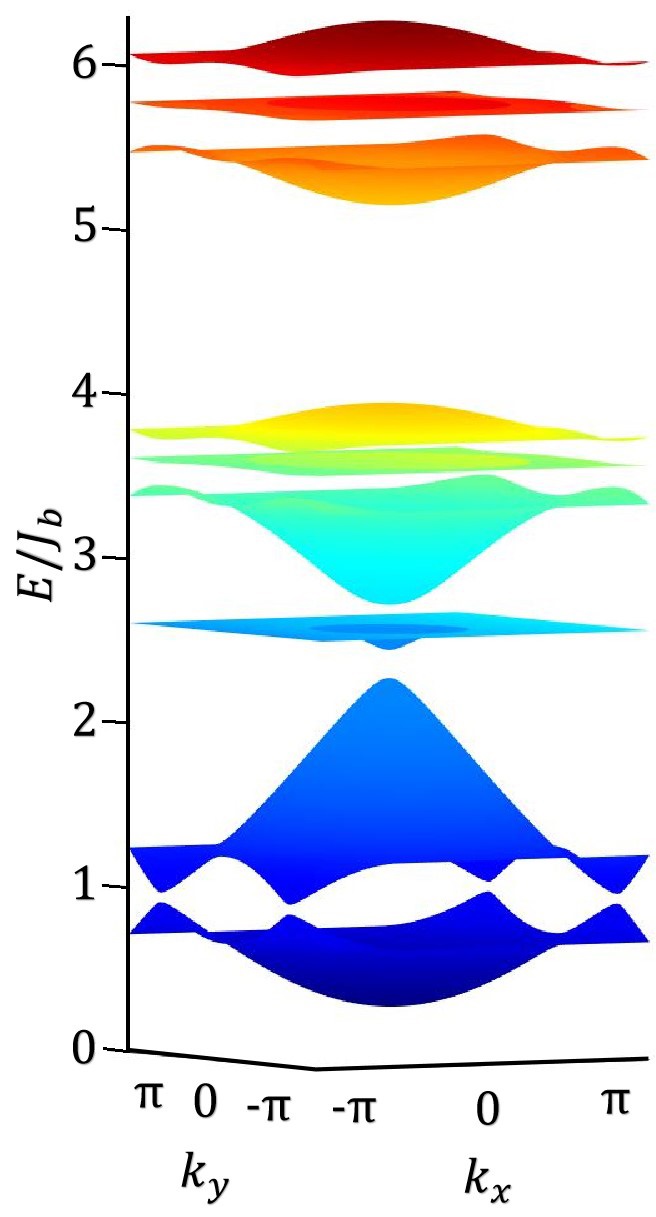}\label{fig2b}}
}
{
\subfigure[]{
\includegraphics[width=1.0in]{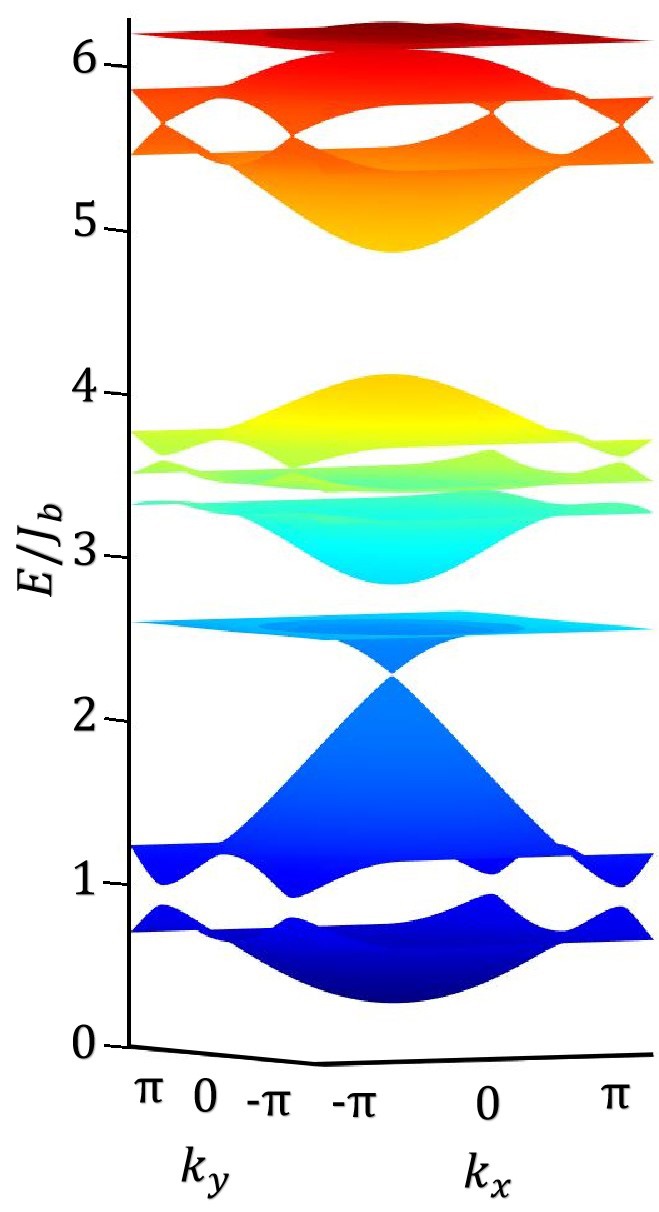}\label{fig2c}}
}
{
\subfigure[]{
\includegraphics[width=1.0in]{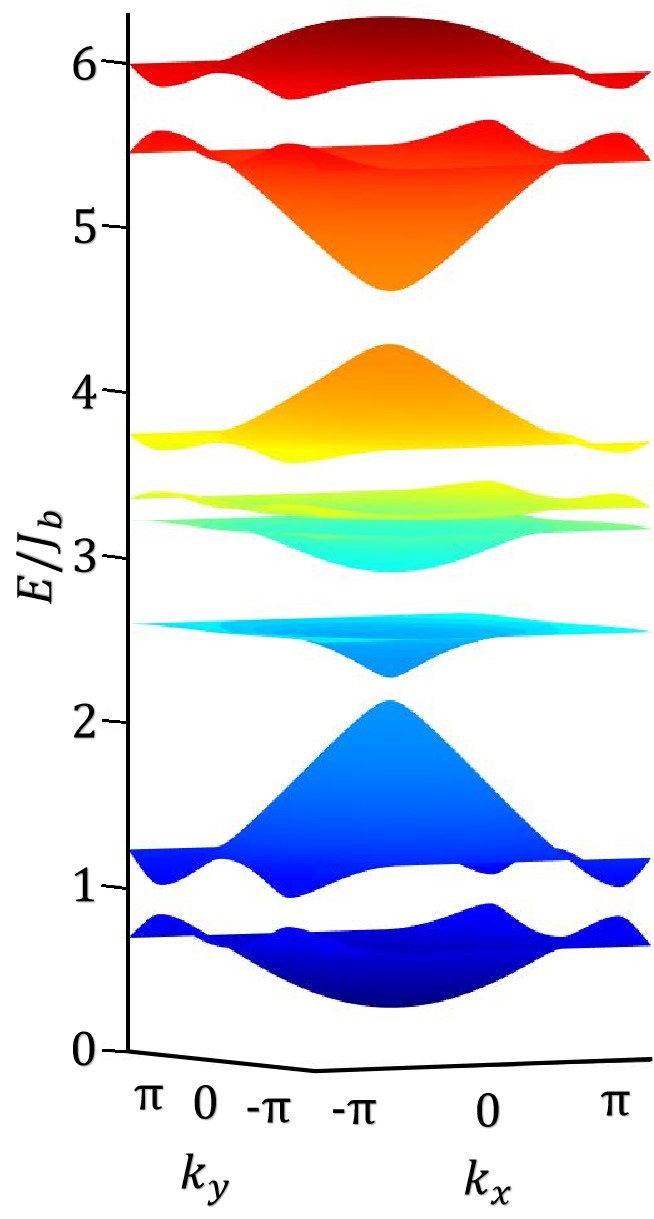}\label{fig2d}}
}
{
\subfigure[]{
\includegraphics[width=1.0in]{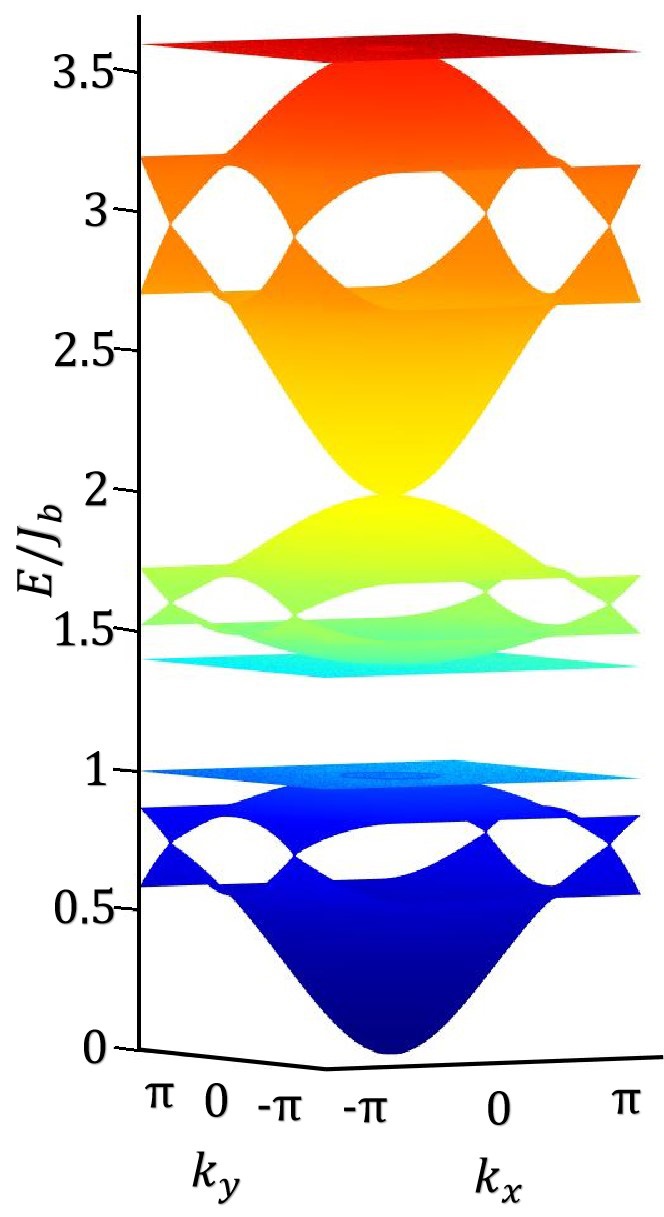}\label{fig2e}}
}
{
\subfigure[]{
\includegraphics[width=1.0in]{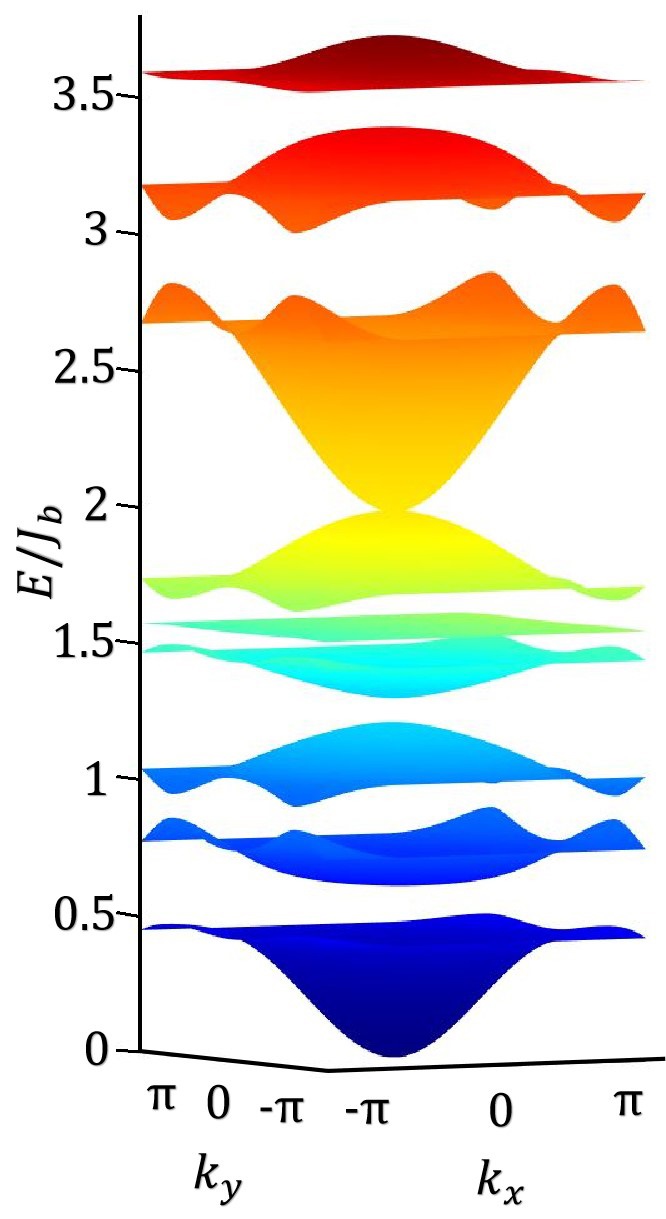}\label{fig2f}}
}
\caption{The magnon bands of the TKL with various DM interactions. The Dirac points are located at K point ($\frac{2}{3}\pi$, $-\frac{2}{3}\pi$) and K' point ($-\frac{2}{3}\pi$, $\frac{2}{3}\pi$) in the first Brillouin zone. The parameters of ferromagnetic coupling with $K$=0.1, $h$=0.1: (a) $J_{a}$=0.5, $J_{b}$=1, $D$=0. (b) $J_{a}$=0.5, $J_{b}$=1, $D$=0.1. (c) $J_{a}$=0.5, $J_{b}$=1, $D$= 0.2. (d) $J_{a}$=0.5, $J_{b}$=1, $D$=0.3. The parameters of ferrimagnetic ground state with $K$=0.1, $h$=0.1 are set as: (e) $J_{a}$=0.2, $J_{b}$=-1, $D$=0. (f) $J_{a}$=0.2, $J_{b}$=-1, $D$=0.1.}
\label{Fig2}
\end{figure*}
%%%%%%%%%%%%%%%%%%%%%%%%%%%%%%%%%%%%%%%%%%%%%%%%%%%%%%%%%%%%%%%%
\subsection{Angular Momentum and Gyromagnetic Ratio}\label{subsec:AMGR}
%%%%%%%%%%%%%%%%%%%%%%%%%%%%%%%%%%%%%%%%%%%%%%%%%%%%%%%%%%%%%%%%
There are correction terms to the thermal Hall conductivity in the linear response theory, by noting that the temperature gradient is not a dynamical force but a statistical force. Thus, the transport coefficients for magnons consist of the deviations of a particle density operator and the current operators. The current operators are expressed in terms of the reduced orbital angular momentum of magnons.
\begin{align}
&l_{edge} = \frac{2k_B}{4\pi^2 \hbar} 2\mathrm{Im} \sum_{n,\boldsymbol{k}} \big \langle \frac{\partial \psi_{n}} {\partial k_x} |Tc_1 (\rho(\varepsilon_{n \boldsymbol{k}})) -\frac{\rho(\varepsilon_{n \boldsymbol{k}}) \varepsilon_{n \boldsymbol{k}} }{k_B}|\frac{\partial\psi_{n}}{\partial k_y}\big \rangle,\non\\
\end{align}
where $c_1(x)$=$(1+x)\ln(1+x)$-$x\ln x$ is another weight function. Besides, the magnon wave packet carries an additional self-rotation motion originating from Berry curvature~\cite{physrevb.84.184406,physrevlett.106.197202}
\begin{align}
&l_{self} = \frac{2k_B}{4\pi^2 \hbar} 2\mathrm{Im} \sum_{n,\boldsymbol{k}} \big \langle \frac{\partial\psi_{n}}{\partial k_x} |\frac{\rho(\varepsilon_{n \boldsymbol{k}})} {2k_B} (\varepsilon_{n \boldsymbol{k}} -H)|\frac{\partial\psi_{n}}{\partial k_y}\big \rangle.
\end{align}

We calculate the total angular momentum per unit cell by summing the edge current and the self-rotation.
\begin{align}
L^{}_{tot} = m^{\ast}(l^{}_{edge}+l^{}_{self}),
\end{align}
where $L_{tot}$ represents the total angular momentum. Within the low temperature approximation, the mass of the magnon can be approximated as the effective mass $m^{\ast}$ at the $\Gamma$ point of the first band. Thus, the gyromagnetic ratio of magnons can be expressed as
\begin{align}
\gamma_m=\frac{\gamma_e L_{tot}}{\hbar \Delta m},
\end{align}
where the $\gamma_e$ is given by $2m_e/(ge)$, $g$ is the Lande factor, $e$ and $m_e$ are the charge and mass of the electron, respectively. Then we define a differential gyromagnetic ratio response $\gamma_m^{\ast}$ as
\begin{align}
\gamma_m^{\ast}=(\frac{\partial L_{tot}/\partial T}{\partial \Delta m/\partial T})_{h},
\end{align}

Different from the electron systems, the gyromagnetic ratio response of topological magnons cannot be measured simply in experiment, but from a response to a temperature change.
%%%%%%%%%%%%%%%%%%%%%%%%%%%%%%%%%%%%%%%%%%%%%%%%%%%%%%%%%%%%%%%%
\section{RESULTS}\label{sec:results}
%%%%%%%%%%%%%%%%%%%%%%%%%%%%%%%%%%%%%%%%%%%%%%%%%%%%%%%%%%%%%%%
\subsection{Topological Magnon Bands}\label{subsec:teb}
%%%%%%%%%%%%%%%%%%%%%%%%%%%%%%%%%%%%%%%%%%%%%%%%%%%%%%%%%%%%%%%
\begin{figure}[t]
\centering
{
\subfigure[]{
\includegraphics[width=3.1in]{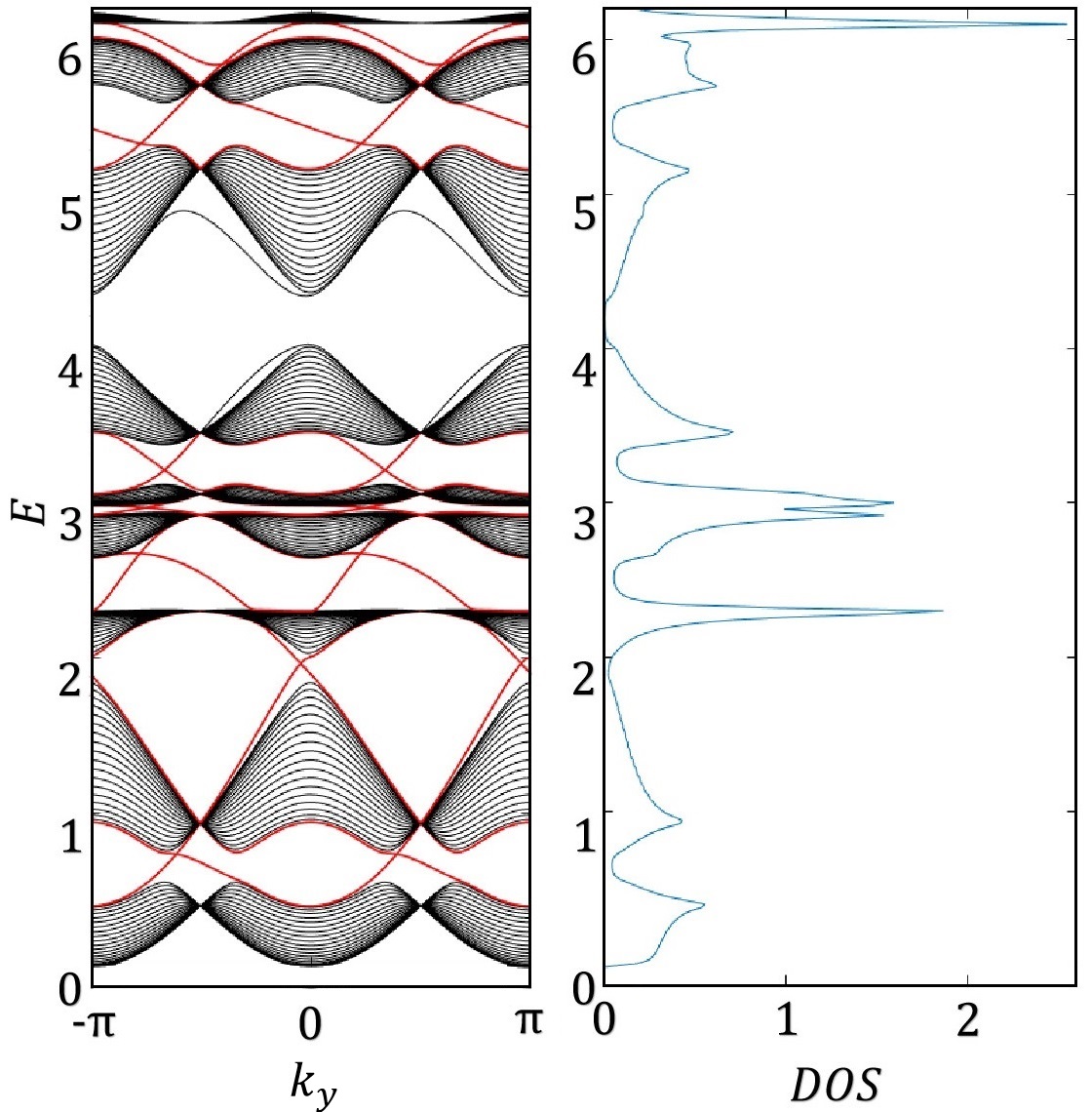}\label{fig3a}}
}
{
\subfigure[]{
\includegraphics[width=3.1in]{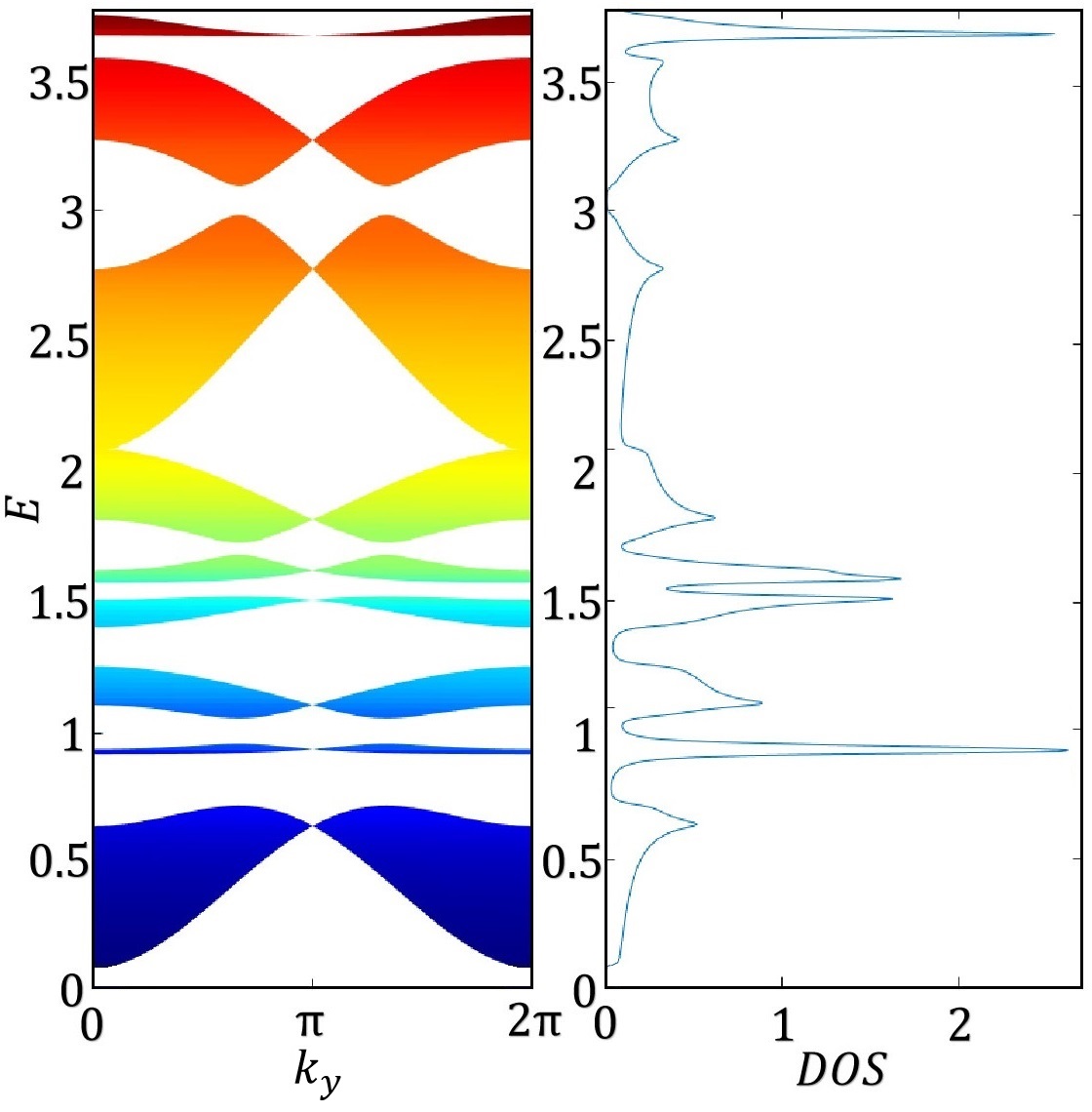}\label{fig3b}}
}
\caption{Magnon density of states with energy bands on the TKL. (a) The magnon band structure and DOS corresponding to topological edge states for a TKL ribbon ($J_{a}$=0.5, $J_{b}$=1, $K$=0.1, $h$=0.1 and $D$=0.3). The dispersions of the edge states in gaps are shown by red curves. (b) The magnon band structure and DOS of the TKL in ferrimagnetic state with $J_{a}$=0.2, $J_{b}$=-1, $K$=0.1, $h$=0.1 and $D$=0.05.}
\label{Fig3}
\end{figure}

Here we target on the ferrromagnetic and ferrimagnetic ground states of the TKL. As shown in Fig.~\ref{Fig2}, the DM interaction which breaks the time reversal symmetry can open the gap at the Dirac points. Thus, we study the topological magnon bands on the TKL and take $|J_{b}|$ as the unit of energy while $J_{a}$=0.5, $J_{b}$=1, $K$=0.1, and $h$=0.1. For ferromagnetic $J_{b}$, we consider the DM value at $D$ = 0, 0.1, 0.2, 0.3 while the numerical solutions of the energies at the high symmetry point $\Gamma$ are given in Table~\ref{tab01}.
\begin{table}[htbp]
\centering
\caption{Energy of each band at $\Gamma$ point with $J_{a}$=0.5, $J_{b}$=1.}
\label{tab01}
\begin{tabular}{cccc ccccc}%表格中的数据居中，c 的个数为表格的列数
\hline\hline\noalign{\smallskip}	
$D$ & Energy (from lower to higher) \\
\noalign{\smallskip}\hline\noalign{\smallskip}
0 & $\{0.30, 2.30, 2.61, 2.61, 3.80, 3.80, 5.49, 5.49, 6.30\}$ \\
0.1 & $\{0.30, 2.30, 2.47, 2.75, 3.63, 3.97, 5.18, 5.80, 6.30\}$ \\
0.2 & $\{0.30, 2.30, 2.32, 2.86, 3.45, 4.15, 4.89, 6.13, 6.30\}$ \\
0.3 & $\{0.30, 2.16, 2.30, 2.94, 3.28, 4.32, 4.64, 6.30, 6.53\}$ \\
\noalign{\smallskip}\hline
\end{tabular}
\end{table}

The Dirac points are located at K point ($\frac{2}{3}\pi$, $-\frac{2}{3}\pi$) and K' point ($-\frac{2}{3}\pi$, $\frac{2}{3}\pi$) in the first Brillouin zone. Hence, we also calculate the numerical solutions of high symmetry point K in Table~\ref{tab02} while K' is equivalent. Additionally, for $J_{a}$=1, $J_{b}$=1 the top band becomes threefold degenerate. As an analog of a spin-orbit interaction in electronic topological insulators, DM interactions can introduce nonzero Berry curvature and change the Chern numbers of some magnon bands. In Table~\ref{tab03}, we numerically check the Chern numbers of different magnon bands, which can distinguish various topological phases.
\begin{table}[htbp]
\centering
\caption{Energy of each band at K point with $J_{a}$=0.5, $J_{b}$=1.}
\label{tab02}
\begin{tabular}{cccc ccccc}
\hline\hline\noalign{\smallskip}	
$D$ & Energy (from lower to higher) \\
\noalign{\smallskip}\hline\noalign{\smallskip}
0 & $\{0.96, 0.96, 2.62, 3.55, 3.55, 3.80, 5.49, 5.89, 5.89\}$ \\
0.1 & $\{0.93, 0.99, 2.61, 3.47, 3.61, 3.75, 5.54, 5.77, 6.04\}$ \\
0.2 & $\{0.90, 1.01, 2.61, 3.36, 3.61, 3.65, 5.68, 5.68, 6.20\}$ \\
0.3 & $\{0.86, 1.04, 2.61, 3.24, 3.42, 3.67, 5.61, 5.87, 6.29\}$ \\
\noalign{\smallskip}\hline
\end{tabular}
\end{table}

\begin{table}[htbp]
\centering
\caption{The Chern numbers with $J_{a}$=0.5, $J_{b}$=1.}
\label{tab03}
\begin{tabular}{cccc ccccc}
\hline\hline\noalign{\smallskip}	
$D$ & the Chern number of each Band (from lower to higher) \\
\noalign{\smallskip}\hline\noalign{\smallskip}
0 & $\{0, 0, 0, 0, 0, 0, 0, 0, 0\}$ \\
0.1 & $\{-1, 1, -1, 2, 0, -1, 1, 0, -1\}$ \\
0.2 & $\{-1, 1, -1, 2, -2, 1, 0, 1, -1\}$ \\
0.3 & $\{-1, 0, 0, 2, -2, 1, -1, 0, 1\}$ \\
\noalign{\smallskip}\hline
\end{tabular}
\end{table}

If $J_a$$>$0 and $J_b$$<$0, the ground state is the ferrimagnetic state and a small DM interaction can change the band structure significantly. From our calculations the antiferromagnetic coupling is unfavourable for energy band topology. Here we choose the DM value at $D$ = 0, 0.1 for antiferromagnetic coupling with $J_{a}$=0.2, $J_{b}$=-1, $K$=0.1, and $h$=0.1. In this case, the bands resemble three copies of magnon bands on the kagome lattice ferromagnet with a flat band and three dispersive Dirac magnon bands in each copy as shown in Fig.~\ref{fig2e} and Fig.~\ref{fig2f}~\cite{PhysRevB.89.134409}. For nonzero DM interaction, the magnon bands are separated by a finite energy gap proportional to the DM interaction in all the parameter regions~\cite{PhysRevB.98.094419,PhysRevB.97.134411}. And the Chern numbers are shown in Table~\ref{tab04}.
\begin{table}[htbp]
\centering
\caption{The Chern numbers with $J_{a}$=0.2, $J_{b}$=-1.}
\label{tab04}
\begin{tabular}{cccc ccc}
\hline\hline\noalign{\smallskip}	
$D$ & The Chern number of each band (from lower to higher) \\
\noalign{\smallskip}\hline\noalign{\smallskip}
0 & $\{0, 0, 0, 0, 0, 0, 0, 0, 0\}$ \\
0.1 & $\{1, 0, -1, -1, 2, -1, -1, 0, 1\}$ \\
\noalign{\smallskip}\hline
\end{tabular}
\end{table}

The Berry curvature and the Chern number can be positive or negative. Both of them become zero when we adjust some of the nine bands to topologically trivial phases, and the summation of Chern numbers for all bands is always zero.

%%%%%%%%%%%%%%%%%%%%%%%%%%%%%%%%%%%%%%%%%%%%%%%%%%%%%%%%%%%%%%%
\subsection{ Armchair Edge States}\label{subsec:dos}
%%%%%%%%%%%%%%%%%%%%%%%%%%%%%%%%%%%%%%%%%%%%%%%%%%%%%%%%%%%%%%%
According to the bulk-edge correspondence, the summation of Chern numbers up to the $j$-th band is equal to the number of pairs of edge states in the gap. We calculate the bulk-edge energy spectrum which corresponds to the surface property of the ribbon sample. The gapless edge states and the DOS are shown in Fig.~\ref{Fig3}. We choose a $9\times20$ lattice and introduce the Green's functions to calculate the armchair edge states of our ribbon sample. The emerging peaks of DOS are dependent on the topological band structure. We expect to derive the value of Chern number for each distinct band from the edge state pattern itself.

As a result, the dispersion of armchair edge states in one-dimensional Brillouin zone is shown in fig.~\ref{Fig3} with $J_a$=0.5, $J_b$=1, $K$=0.1, $h$=0.1 and $D$=0.3. We also calculate the DOS of a two-dimensional TKL system with ferrimagnetic ground state for $J_{a}$=0.2, $J_{b}$=-1, $K$=0.1, $h$=0.1 and $D$=0.05. It can be written as a sum of Dirac-$\delta$ functions with energies corresponding to the set of eigenvalues of the Hamiltonian. The appearance of edge modes leads to nonzero DOS in each DM-induced gap. And the DOS is no longer symmetric about the Dirac point. The topological structure of energy bands is described by the magnon transport of armchair edge states and the corresponding DOS.
%%%%%%%%%%%%%%%%%%%%%%%%%%%%%%%%%%%%%%%%%%%%%%%%%%%%%%%%%%%%%%%
\subsection{ Thermal Hall Effect}\label{subsec:mhe}
%%%%%%%%%%%%%%%%%%%%%%%%%%%%%%%%%%%%%%%%%%%%%%%%%%%%%%%%%%%%%%%
Thermal Hall effect is a key experimental signature to detect the magnon transport arise from the edge current of topological excitations. The DM-induced Berry curvature acts as an effective magnetic field that deflects the propagation of magnon in the system. The nonzero Chern numbers are associated with topological chiral gapless edge modes which appear in the DM-induced gaps. And the nontrivial topology of the Berry curvature leads to magnon edge states which carry a transverse heat current upon the application of a longitudinal temperature gradient. Unlike electrons the magnons have no charge and the rotation is not due to "Lorentz force". Thus, the DM interaction plays a role of an effective magnetic field by altering the propagation of magnons in the system~\cite{physrevlett.115.106603}.

\begin{figure}[t]
\centering
{
\includegraphics[width=3.4in]{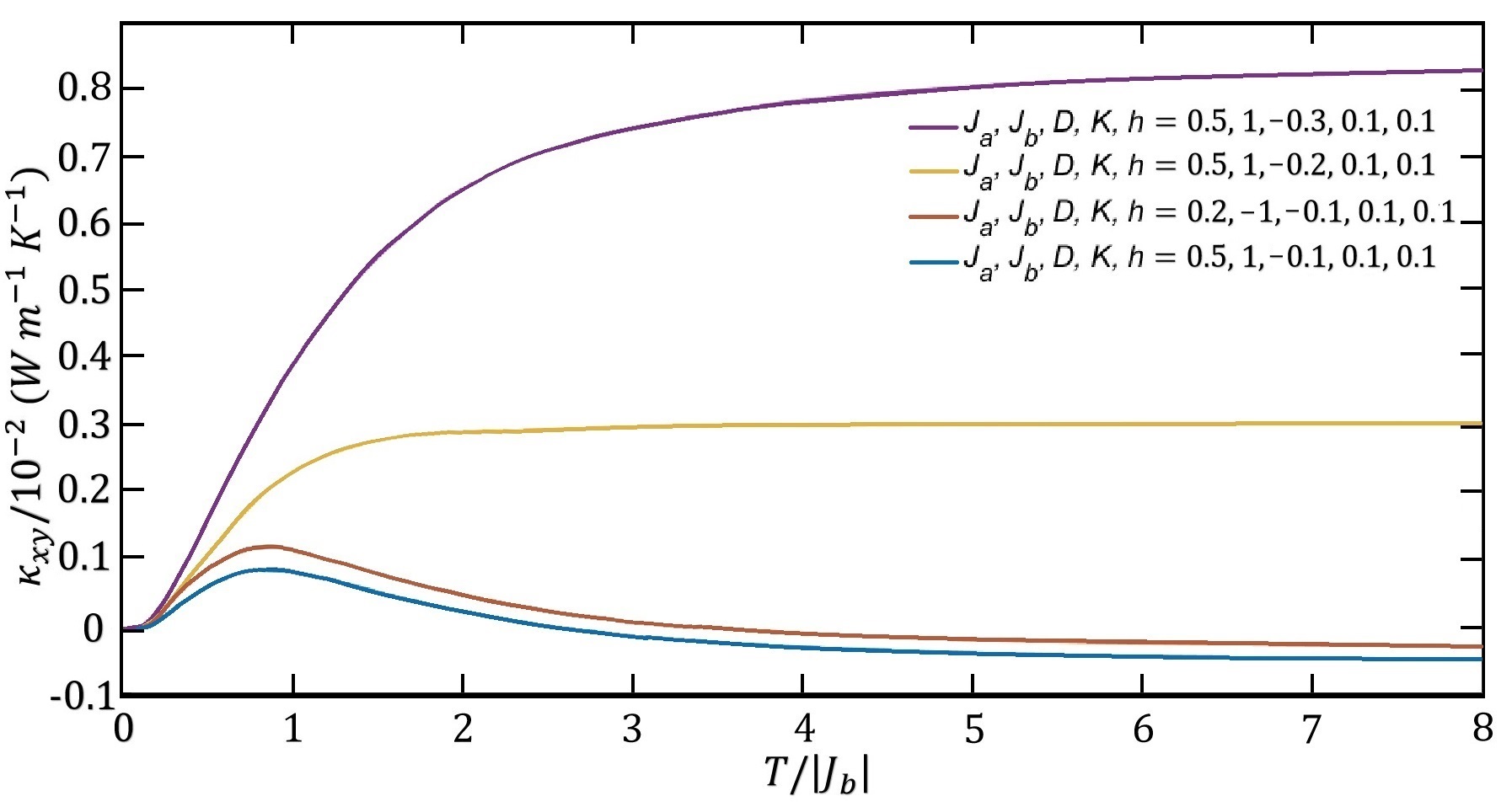}\label{fig5}
}

\caption{Low-temperature thermal Hall conductivity on the TKL. In ferromagnetic coupling the DM interactions are $-0.1, -0.2, -0.3$ with $J_{a}$=0.5, $J_{b}$=1, $K$=0.1, $h$=0.1 respectively. For antiferromagnetical couplings between sublattices $A$ and $B$, the DM interaction is \textcolor[rgb]{1,0,0}{$-0.1$} with $J_{a}$=0.2, $J_{b}$=-1, $K$=0.1, $h$=0.1.}
\label{Fig5}
\end{figure}

The plot of $\kappa^{xy}(T)$ vs. $T/|J_b|$ is displayed with different DM interactions respectively in Fig.~\ref{Fig5}. We take $h$=0.1 and $K$=0.1 for both ferromagnetic state and ferrimagnetic state. Because of the opposite Berry curvatures of the higher magnon bands, the $\kappa^{xy}(T)$ for $D$=-0.1 changes its sign upon raising the temperature. We observe that the thermal Hall conductivity decreases with the emergence of antiferromagnetic coupling. Especially when the magnon is excited to the energy band possessing a high Chern number, the $\kappa^{xy}$ can be effectively changed with a positive peak at low temperatures followed by a long negative tail in the high-temperature region.

As shown in Fig.~\ref{Fig5}, we also calculate the thermal Hall conductivity coefficients from HP theory with $D$=-0.2, -0.3. The values keep increasing upon raising the temperature and have not reached saturation at the phase transition points. At low temperature and weak field, the lowest-lying magnon band dominates thermal transport. The thermal Hall conductivity vanishes at zero temperature as there are no thermal excitations and becomes negative ($\kappa^{xy}<$0) due to the fact that $c_{2}(x)>$0. At high temperature, higher-energy bands carrying opposite Berry flux contribute significantly and a strong Zeeman field diminishes the thermal population difference among the bands by creating a large gap for all the bands. These behaviors of $\kappa^{xy}$ are inherited from the topology of the bulk magnon bands~\cite{physrevlett.115.106603,physrevb.91.125413}. The peak values and convergence values for the different parameters on the TKL are listed in Table~\ref{tab05}.

\begin{table}[htbp]
\centering
\caption{The peak and convergence values for different parameters.}
\label{tab05}
\begin{tabular}{cccc ccccc}%表格中的数据居中，c 的个数为表格的列数
\hline\hline\noalign{\smallskip}	
Parameters &  Peak value & Convergence value  \\
\noalign{\smallskip}\hline\noalign{\smallskip}
$J_a$=0.2, $J_b$=-1, $D$=-0.1 & 0.124 & -0.030 \\
$J_a$=0.5, $J_b$=1, $D$=-0.1 & 0.091 & -0.050 \\
$J_a$=0.5, $J_b$=1, $D$=-0.2 & \ & 0.301 \\
$J_a$=0.5, $J_b$=1, $D$=-0.3 & \ & 0.832 \\
\noalign{\smallskip}\hline
\end{tabular}
\end{table}

It is noted that the energy bands with high Chern numbers have large weights in the calculation of thermal Hall effect and the dominant contribution comes from the Dirac points $K$ ($K'$). With the enhancement of DM interaction, the bands with high Chern numbers appear. Thus, both of the sign change and the peak vanish in all the parameter regions and $\kappa^{xy}$ increases significantly. In real materials, the Curie temperature can be increased significantly  due to the presence of single-ion anisotropies, interlayer couplings, and so on. Here we show the results of thermal Hall conductivity on the TKL in a large temperature range to illustrate the behaviors of $\kappa^{xy}$ for different parameters.

\begin{figure}[t]
\centering
{
\includegraphics[width=3.4in]{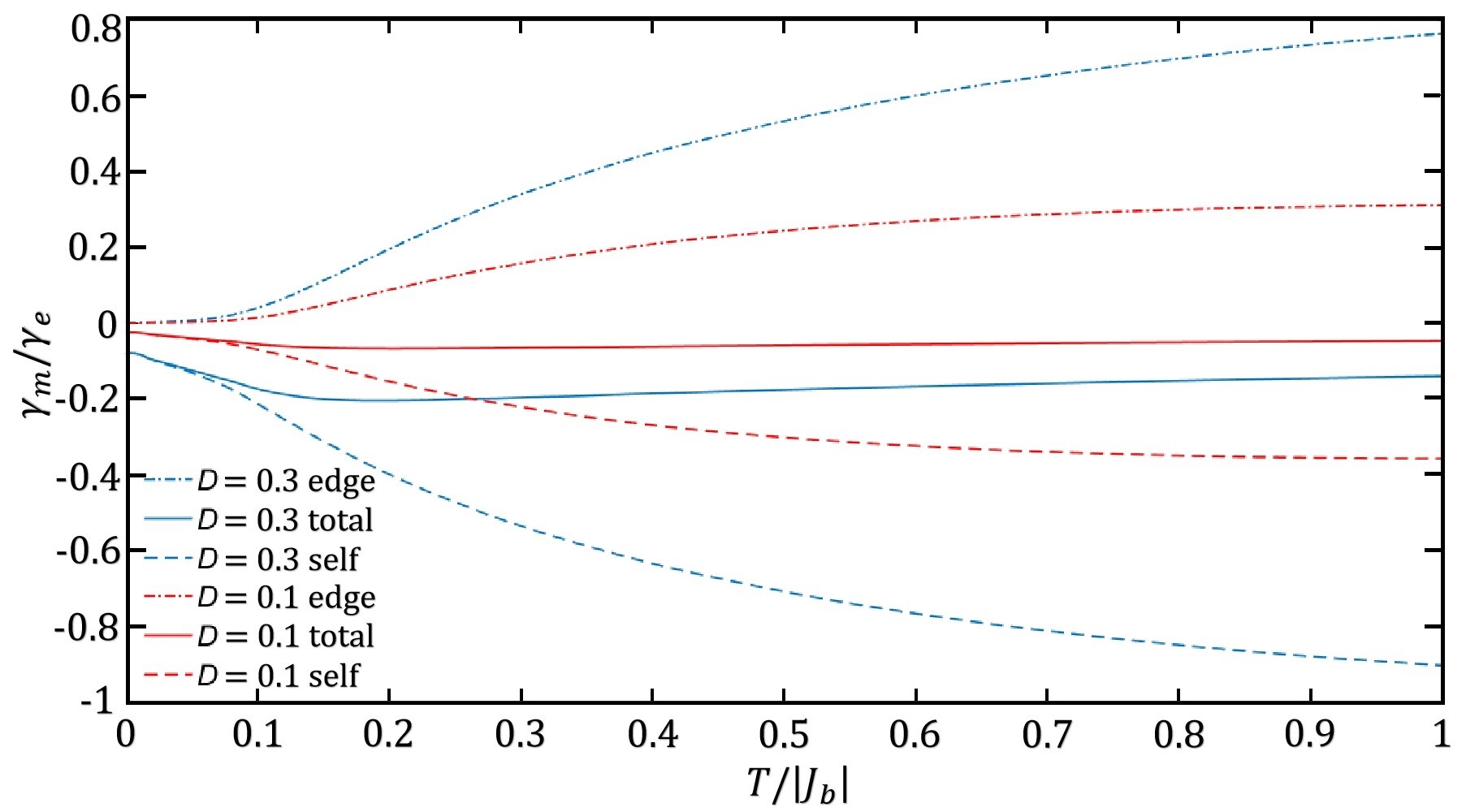}\label{fig6}
}

\caption{Temperature dependence of the topological gyromagnetic ratio $\gamma_m$ on the TKL. The gyromagnetic ratio contributions, compared to the electronic value, for individual topological edge current, self-rotation of the magnon wave packet, and the total angular momentum contribution to $\gamma_m$ are shown. Parameters choices are $D$=0.1, 0.3 with $J_{a}$=0.5, $J_{b}$=1, $K$=0.1, and $h$=0.1.}
\label{Fig6}
\end{figure}
%%%%%%%%%%%%%%%%%%%%%%%%%%%%%%%%%%%%%%%%%%%%%%%%%%%%%%%%%%%%%%%%
\subsection{ Einstein-de Haas Effect}\label{subsec:EdH}
%%%%%%%%%%%%%%%%%%%%%%%%%%%%%%%%%%%%%%%%%%%%%%%%%%%%%%%%%%%%%%%
\begin{figure*}[t]
\centering
{
\subfigure[]{
\includegraphics[width=3.4in]{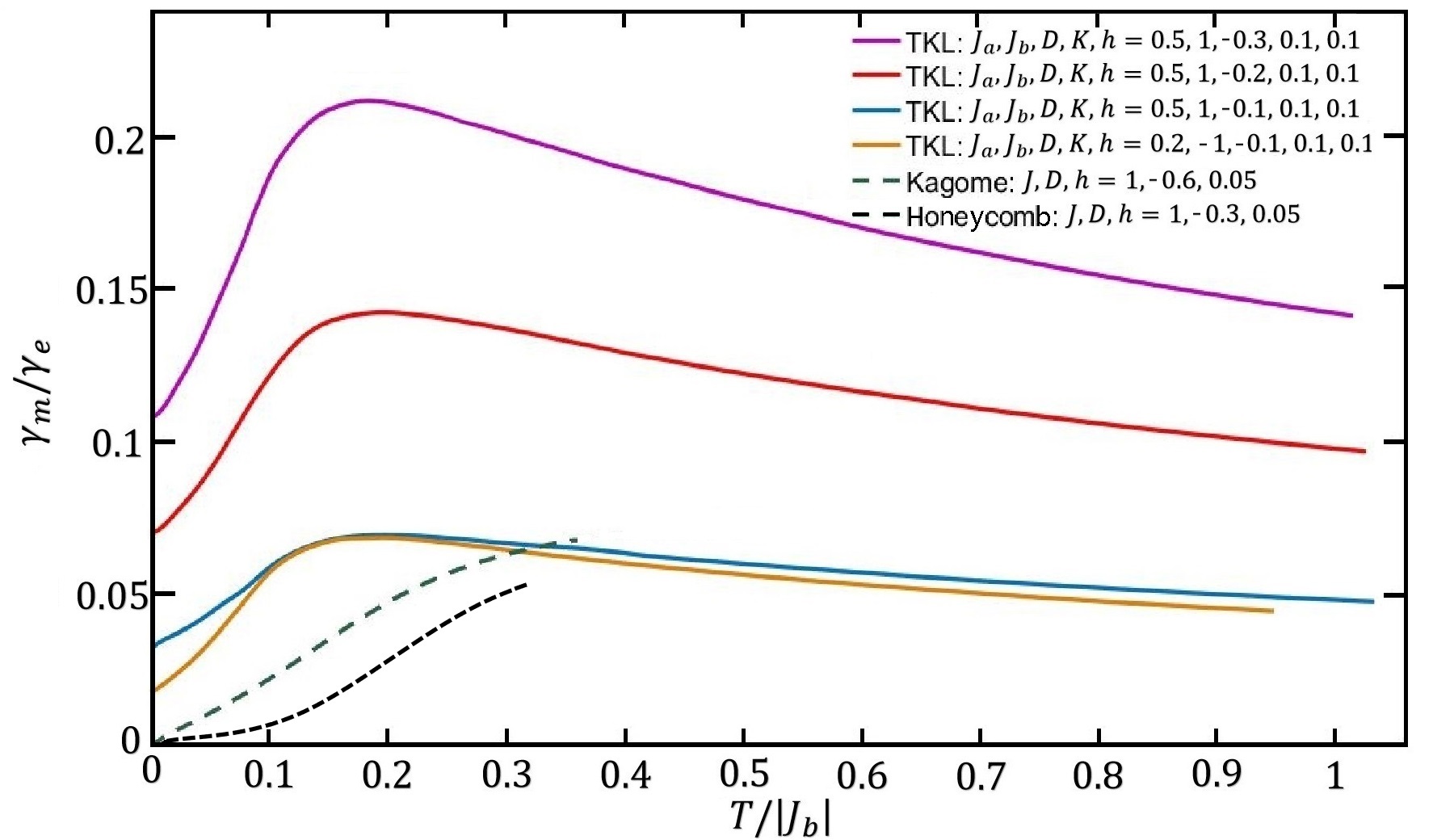}\label{fig7a}}
}
{
\subfigure[]{
\includegraphics[width=3.4in]{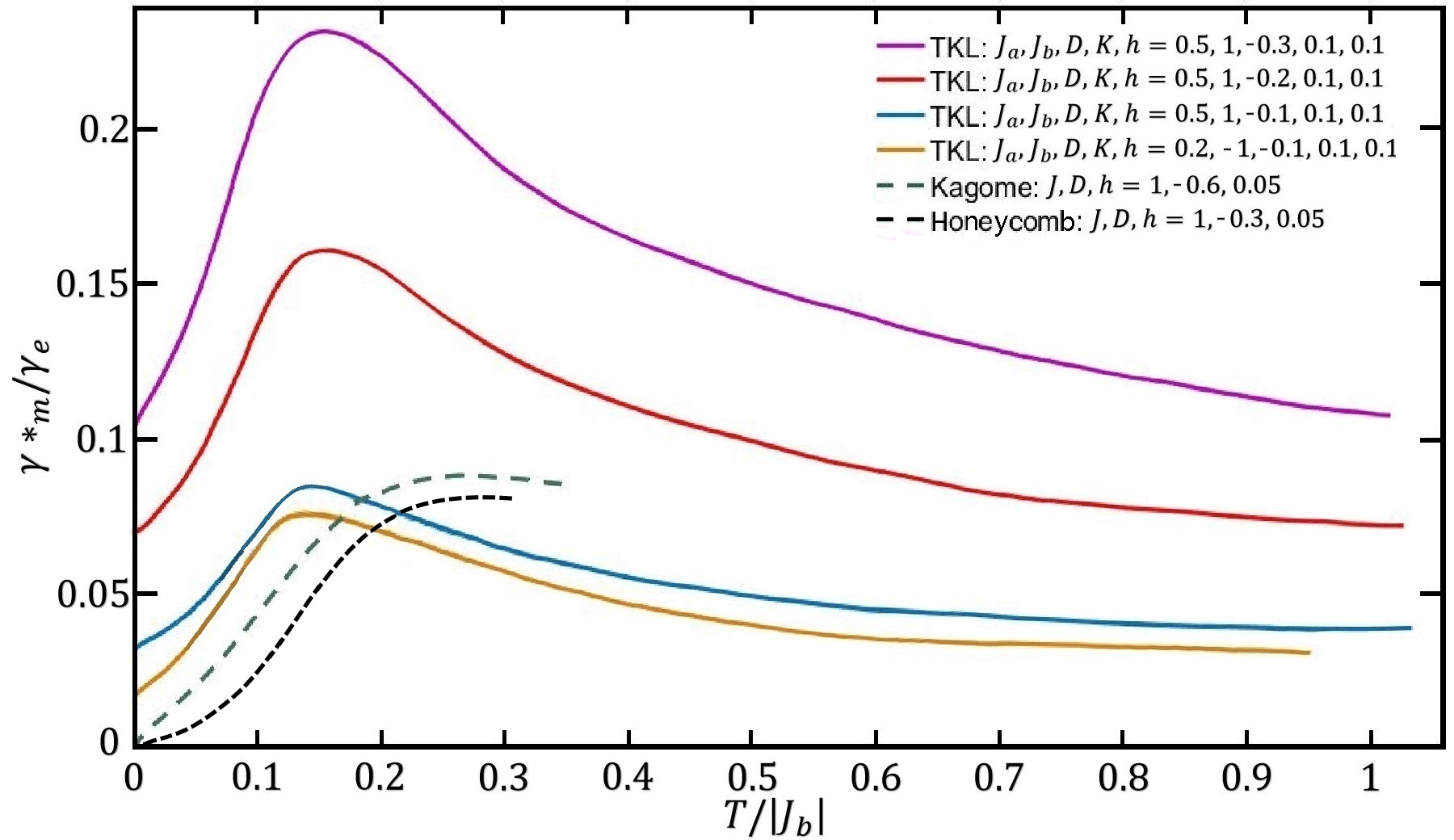}\label{fig7b}}
}
\caption{Comparison of the Einstein de-Haas effect response on the TKL and kagome lattice. (a) The gyromagnetic ratio variation with temperature is shown. Parameters choices on the kagome lattice are $J$=$J_{b}$=1, $D$=-0.1 and $h$=0.3. Parameters choices on the TKL with $K$=0.1 and $h$=0.1 are $D$=-0.1, -0.2, -0.3 for $J_{a}$=0.5, $J_{b}$=1 and $D$=-0.1 for $J_{a}$=0.2, $J_{b}$=-1. (b) The differential gyromagnetic ratio response is shown. Parameter choices are the same as before.}
\label{Fig7}
\end{figure*}
The Einstein-de Haas effect is the ultimate macroscopic manifestation originating from a subtle microscopic exchange of spin angular momentum~\cite{PhysRevLett.112.085503}. According to the linear response theory, the magnon wave packet undergoes two types of orbital motions and the total angular momentum is defined as the summation of these two types of rotational motions. We define the gyromagnetic ratio as the angular momentum divided by the magnetic moment of magnons, which is related to the magnetization change of the system. Each magnon mode can be excited or annihilated and has its own gyromagnetic response. In Fig.~\ref{Fig6}, we calculate the transport properties from the gyromagnetic ratio, finding that the self-rotation motion and the edge current on the TKL are opposite in directions. However, the self-rotation has a larger part in the negative region, which results in a negative value of the total angular momentum in all temperature region. Especially in the zero temperature limit, the self-rotation motion of the TKL has a finite response at the $\Gamma$ point which sets it apart from the other usual lattice candidates that have been explored before.

Our results refer that the total gyromagnetic contribution increases significantly at first and reaches a peak value at about $T$=0.20$|J_b|$. The value stabilizes when it approaches the Curie temperature $T_c$. Since the $T_c$ is in the ballpark of $|J_{b}|$, the HP representation is valid until $T$ $\leq$ $T^{*}$ ($T^{*}$ $\sim$ 0.5$|J_{b}|$), providing a quantitative description of the thermal Hall conductivity and the gyromagnetic ratio in the temperature range [0, $T^{*}$]. Above $T^{*}$, the results obtained from the HP representation illustrate the trends that one would expect from a more accurate calculation. Since the relevant topological features of the EdH response happen in the ballpark of $T$ = 0.2$|J_{b}|$, the HP representation is enough to describe them. The Curie temperature for the TKL are listed in the Table~\ref{tab06}. To further analyze the physical content of Fig.~\ref{Fig6}, we compare the gyromagnetic ratio of the TKL system with respect to the kagome lattice system. We show the results of our calculation in Fig.~\ref{Fig7}. The $\gamma_m/\gamma_e$ shown in Fig.~\ref{fig7a} represents the temperature variation of the topological gyromagnetic ratio compared to the electronic value. As the $\gamma_e^{\ast}$ is equal to $\gamma_e$ for electrons, the $\gamma_m^{\ast}/\gamma_e^{\ast}$ can be simplified as $\gamma_m^{\ast}/\gamma_e$. Hence, the differential gyromagnetic ratio is renormalized from the $\gamma_m$ response. From our calculations we find that the ferrimagnetic frustrated structure suppresses the band topology by reducing the $\gamma_m/\gamma_e$ and $\gamma_m^{\ast}/\gamma_e$. Considering the differential gyromagnetic ratio response, the magnon system also has a peak value before descending as seen in Fig.~\ref{fig7b}. Thus, there is an optimal temperature of the differential gyromagnetic ratio at which the magnon insulator will have the strongest response.

\begin{table}[htbp]
\centering
\caption{The Curie temperature $T_c$/$|J_b|$ for different parameters.}
\label{tab06}
\begin{tabular}{cccc ccccc}%表格中的数据居中，c 的个数为表格的列数
\hline\hline\noalign{\smallskip}	
Lattice &  Parameters field & $T_c$/$|J_b|$  \\
\noalign{\smallskip}\hline\noalign{\smallskip}
TKL & $J_a$=0.5, $J_b$=1, $D$=-0.1, $K$=0.1 & 0.884 \\
TKL & $J_a$=0.5, $J_b$=1, $D$=-0.2, $K$=0.1 & 0.878 \\
TKL & $J_a$=0.5, $J_b$=1, $D$=-0.3, $K$=0.1 & 0.867 \\
TKL & $J_a$=0.2, $J_b$=-1, $D$=-0.1, $K$=0.1 & 0.816 \\
\noalign{\smallskip}\hline
\end{tabular}
\end{table}

In Stewart’s apparatu, the EdH effect was observed from the amount of the transient angular momentum change. This experimental setup can be explored to measure the differential gyromagnetic ratio via being exposed to an external heat bath with a temperature gradient~\cite{PhysRev.11.100}. From an experimental point of view, there is an optimal temperature zone in which our theory can be tested well. The values of optimal temperature for various TKL systems are all around $T=0.20|J_b|$. Additionally, the anisotropy and external magnetic field can enhance the EdH effect, for instance, the peak of $\gamma_m^{\ast}/\gamma_e$ reaches 0.223 for the ferromagnetic TKL with $D$=-0.3, $J_{a}$=0.5, $J_{b}$=1, $K$=0.1 and $h$=0.1. Our formalism, analytical approach, and eventual conclusions will hold not only for the TKL system, but also for a wider variety of ferrimagnetic systems.
%%%%%%%%%%%%%%%%%%%%%%%%%%%%%%%%%%%%%%%%%%%%%%%%%%%%%%%%%%%%%%%
\section{CONCLUSIONS}\label{sec:conclu}
%%%%%%%%%%%%%%%%%%%%%%%%%%%%%%%%%%%%%%%%%%%%%%%%%%%%%%%%%%%%%%%
In summary, we have investigated the topological magnons on the TKL, which can give detectable results on the thermal Hall conductance and the Einstein-de Haas effect. In the presence of armchair edges for a ribbon sample, we find that the nonzero summation of Chern numbers for different bands below the gap leads to a magnon current transport along the $k_y$ direction of this gap~\cite{PhysRevB.90.024412,PhysRevB.89.134409}. By using the real-space Green's function approach, we have studied the armchair edge modes to calculate the DOS in our sample. Theoretical and experimental studies have shown that thermal Hall conductance can have a sign change as temperature or magnetic field is varied~\cite{PhysRevLett.61.2015}. Our results show that the sign change behaviors emerge on the TKL when the topological features are reduced by the antiferromagnetic coupling. At the Curie temperature, the thermal Hall conductivities are always convergent for all selected parameters. We further find that there is a peak for the thermal Hall conductance when the low magnon bands dominate and the peak vanishes when the DM interaction is strong enough. The influence of a nonzero Berry curvature and its underlying topological identity is preserved even though the lattice structure changes.

We show the calculations for the EdH effect of topological magnons for both the ferro- and ferrimagnetic states and propose that the TKL is a suitable lattice for the observation of the EdH effect. Especially in the low temperature region, the magnon description is more effective. Comparing with the traditional kagome and honeycomb lattices, this compound lattice has a better topological magnon structure with added high Chern numbers to produce stronger EdH effect. We investigate the angular momentum for topological edge current and self-rotation originating from the Berry curvature in momentum space. These two angular momentum components with opposite signs offset each other, but the self-rotation has a larger part which ensures that the total angular momentum contribution has a nonzero value. The EdH effect is a macroscopic mechanical manifestation caused by the angular momentum conservation principle and can be detected by a mechanical experimental setup~\cite{physrevresearch.3.023248,PhysRev.11.100}.

We have studied various TKL systems with different coupling parameters to explore the topological magnon excitations, the thermal Hall effect~\cite{Zhang_2016} and the EdH effect. In real materials, the observed results are influenced by other kinds of effects, but these may not be a concern since systems in which topological magnons already dominate the thermal Hall effect and the EdH effect. The TKL structure has been found in Cu$_9$X$_2$(cpa)$_6$(X=F, Cl, Br; cpa=anion of 2-carboxypentonic acid) which has tunable magnetic couplings~\cite{doi:10.1063/1.5130392,PhysRevB.80.132402}. The thermal Hall effect of spin excitations arises in the usual way via the breaking of inversion symmetry of the lattice by a nearest-neighbour DM interaction~\cite{PhysRevLett.105.225901}. It is also possible to realize the TKL in cold atom systems higher-order topology of magnons~\cite{Schindler2018,2019PhRvL.122y6402X}. Our study provides a new vision to realize the thermal Hall effect and the EdH effect. The thermal Hall effect that arises from the edge current of magnons is useful to control the magnon transport, then the EdH effect can produce potential mechanical effect which has potential applications in quantum informatics and topological magnon spintronics~\cite{RevModPhys.90.015005}.

%%%%%%%%%%%%%%%%%%%%%%%%%%%%%%%%%%%%%%%%%%%%%%%%%%%%%%%%%%%%%%%%

%\newpage
\begin{acknowledgments}
We would like to thank Trinanjan Datta and Jun Li for helpful discussions. This project is supported by NKRDPC-2022YFA1402802, NKRDPC-2018YFA0306001, NSFC-11974432, NSFC-92165204, GBABRF-2019A1515011337, Leading Talent Program of Guangdong Special Projects (201626003), and Shenzhen International Quantum Academy (Grant No. SIQA202102).
\end{acknowledgments}

%\newpage
%%%%%%%%%%%%%%%%%%%%% appendix %%%%%%%%%%%%%%%%%%%%%%%%%%%%%%%

%\bibliographystyle{apsrev4-1}
\bibliography{cite}
%\bibliography{pericles_15213749405}
\end{document}